\documentclass{aastex62}

\received{MM DD, YYYY}
\revised{MM DD, YYYY}
\accepted{MM DD, YYYY}
\submitjournal{ApJ}

\shorttitle{Molecular gas in a {\it Spitzer} bubble N4}
\shortauthors{Fujita et al.}

\begin{document}

\title{FUGIN: Molecular gas in a {\it Spitzer} bubble N4: possible evidence for a cloud--cloud collision as a trigger of massive star formations}

\author{Shinji FUJITA}
\affil{Department of Astrophysics, Nagoya University,  \\
Furo-cho, Chikusa-ku, Nagoya, Aichi, Japan 464-8602}

\author{Kazufumi TORII}
\affil{Nobeyama Radio Observatory, National Astronomical Observatory of Japan (NAOJ),   \\
National Institutes of Natural Sciences (NINS), 462-2, Nobeyama,   \\
Minamimaki, Minamisaku, Nagano 384-1305, Japan}

\author{Kengo TACHIHARA}
\affil{Department of Astrophysics, Nagoya University,  \\
Furo-cho, Chikusa-ku, Nagoya, Aichi, Japan 464-8602}

\author{Rei ENOKIYA}
\affil{Department of Astrophysics, Nagoya University,  \\
Furo-cho, Chikusa-ku, Nagoya, Aichi, Japan 464-8602}

\author{Katsuhiro HAYASHI}
\affil{Department of Astrophysics, Nagoya University,  \\
Furo-cho, Chikusa-ku, Nagoya, Aichi, Japan 464-8602}

\author{Nario KUNO}
\affil{Department of Physics, Graduate School of Pure and Applied Sciences,  \\
University of Tsukuba, 1-1-1 Ten-nodai, \\
Tsukuba, Ibaraki, Japan 305-8577}

\author{Mikito KOHNO}
\affil{Department of Astrophysics, Nagoya University,  \\
Furo-cho, Chikusa-ku, Nagoya, Aichi, Japan 464-8602}

\author{Tomoka TOSAKI}
\affil{Department of Geoscience, Joetsu University of Education,   \\
Joetsu, Niigata, Japan 943-8512}

\author{Mitsuyoshi YAMAGISHI}
\affil{Institute of Space and Astronautical Science, \\
Japan Aerospace Exploration Agency, \\
Chuo-ku, Sagamihara 252-5210, Japan}

\author{Atsushi NISHIMURA}
\affil{Department of Astrophysics, Nagoya University,  \\
Furo-cho, Chikusa-ku, Nagoya, Aichi, Japan 464-8602}

\author{Tomofumi UMEMOTO}
\affil{Nobeyama Radio Observatory, National Astronomical Observatory of Japan (NAOJ),   \\
National Institutes of Natural Sciences (NINS), 462-2, Nobeyama,   \\
Minamimaki, Minamisaku, Nagano 384-1305, Japan}

\author{Tetsuhiro MINAMIDANI}
\affil{Nobeyama Radio Observatory, National Astronomical Observatory of Japan (NAOJ),   \\
National Institutes of Natural Sciences (NINS), 462-2, Nobeyama,   \\
Minamimaki, Minamisaku, Nagano 384-1305, Japan}

\author{Mitsuhiro MATSUO}
\affil{Nobeyama Radio Observatory, National Astronomical Observatory of Japan (NAOJ),   \\
National Institutes of Natural Sciences (NINS), 462-2, Nobeyama,   \\
Minamimaki, Minamisaku, Nagano 384-1305, Japan}

\author{Yuya TSUDA}
\affil{Meisei University, 2-1-1 Hodokubo, \\
Hino, Tokyo, Japan 191-0042}

\author{Hidetoshi SANO}
\affil{Department of Astrophysics, Nagoya University,  \\
Furo-cho, Chikusa-ku, Nagoya, Aichi, Japan 464-8602}

\author{Daichi TSUTSUMI}
\affil{Department of Astrophysics, Nagoya University,  \\
Furo-cho, Chikusa-ku, Nagoya, Aichi, Japan 464-8602}

\author{Akio OHAMA}
\affil{Department of Astrophysics, Nagoya University,  \\
Furo-cho, Chikusa-ku, Nagoya, Aichi, Japan 464-8602}

\author{Satoshi YOSHIIKE}
\affil{Department of Astrophysics, Nagoya University,  \\
Furo-cho, Chikusa-ku, Nagoya, Aichi, Japan 464-8602}

\author{Kazuki OKAWA}
\affil{Department of Astrophysics, Nagoya University,  \\
Furo-cho, Chikusa-ku, Nagoya, Aichi, Japan 464-8602}

\author{Yasuo FUKUI}
\affil{Department of Astrophysics, Nagoya University,  \\
Furo-cho, Chikusa-ku, Nagoya, Aichi, Japan 464-8602}

\author{other FUGIN members}



\begin{abstract}

Herein, we present the $^{12}$CO ($J$=1--0) and $^{13}$CO ($J$=1--0) emission line observations via the FOREST Unbiased Galactic plane Imaging survey with the Nobeyama 45-m telescope (FUGIN) toward a {\it Spitzer} bubble N4. 
We observed clouds of three discrete velocities: 16, 19, and 25\,km\,s$^{-1}$.
Their masses were $0.1\times 10^{4}$\,$M_{\odot}$, $0.3\times 10^{4}$\,$M_{\odot}$, and $1.4\times 10^{4}$\,$M_{\odot}$, respectively. 
The distribution of the 25-km\,s$^{-1}$ cloud likely traces the ring-like structure observed at mid-infrared wavelength. 
The 16- and 19-km\,s$^{-1}$ clouds have not been recognized in previous observations of molecular lines. 
We could not find clear expanding motion of the molecular gas in N4. 
On the contrary, we found a bridge feature and a complementary distribution, which are discussed as observational signatures of a cloud--cloud collision, between the 16- and 25-km\,s$^{-1}$ clouds.
We proposed a possible scenario wherein the formation of a massive star in N4 was triggered by a collision between the two clouds. 
The time scale of collision is estimated to be 0.2--0.3\,Myr, which is comparable to the estimated dynamical age of the H{\sc ii} region of $\sim$0.4 Myr. 
In N4W, a star-forming clump located west of N4, we observed molecular outflows from young stellar objects and the observational signature of a cloud-cloud collision. 
Thus, we also proposed a possible scenario in which massive- or intermediate-mass star formation was triggered via a cloud--cloud collision in N4W. 

\end{abstract}

\keywords{ISM: clouds --- ISM: individual objects (N4) --- radio lines: ISM --- stars: formation}


\section{Introduction}

\subsection{Massive star formation}
Massive stars are important objects in the galactic environment due to their powerful influence on the interstellar medium via ultraviolet radiation, stellar winds, and supernova explosions.
It is therefore of fundamental importance to understand the mechanisms of massive star formation, although these mechanisms still remain elusive (e.g., \cite{2007ARA&A..45..481Z, 2014prpl.conf..149T}). 

Recently, supersonic cloud--cloud collision (CCC) has been discussed as an important triggering mechanism of massive star formation because of the large mass accretion associated with the compressed region between colliding two clouds.
Observational studies have suggested the possible evidence of CCCs in H{\sc ii} regions (with one or several young O-stars) and in super star clusters (with 10--20 O-stars) in the Milky Way and the Large Magellanic Cloud (\cite{2014ApJ...780...36F, 2015ApJ...807L...4F, 2018ApJ...859..166F, 2018PASJ...70S..60F, 2018PASJ...70S..46F, 2009ApJ...696L.115F, 2018PASJ...70S..49E, 2018PASJ...70S..48H, 2018PASJ...70S..50K, 2018PASJ...70S..42N, 2017arXiv170606002N, 2010ApJ...709..975O, 2018PASJ...70S..45O, 2018PASJ...70S..47O, 2017arXiv170808149S, 2018PASJ...70S..43S, 2013ApJ...768...72S, 2011ApJ...738...46T, 2015ApJ...806....7T, 2017ApJ...835..142T, 2017arXiv170607164T, 2018PASJ...70S..51T, 2015PASJ...67..109T, 2017arXiv170605664T}). 
Magneto-hydrodynamical simulations of the formation of the massive clumps that may form massive stars in the collision-compressed layer have also been discussed (\cite{2013ApJ...774L..31I, 2018PASJ...70S..53I}). 
Furthermore, comparisons between the observations and numerical calculations (\cite{1992PASJ...44..203H, 2010MNRAS.405.1431A, 2014ApJ...792...63T, 2017arXiv170608656T}) have indicated two important observational signatures of CCCs: the ``bridge feature'' in position-velocity diagrams and the ``complementary spatial distribution'' in the sky between two colliding clouds, which provide useful diagnostics to identify CCCs using molecular line observations (\cite{2018ApJ...859..166F}). 
Bridge features are relatively weak CO emissions at intermediate velocities between two colliding clouds with different velocities.
When a smaller cloud drives into a larger cloud, the smaller cloud caves the larger cloud owing to the momentum conservation (\cite{2015MNRAS.450...10H}), and a dense compressed layer is formed at the collision interface, resulting in a thin, turbulent layer between the larger cloud and the compressed layer. 
If one observes a snapshot of this collision with a viewing angle parallel to the axis of collision, two separated velocity peaks connected by an intermediate--velocity emission with lower intensity feature could be observed in position-velocity diagrams. 


\subsection{{\it Spitzer} bubble N4}
Churchwell et al. (2006, 2007) listed $\sim$600 objects with {\it Spitzer} 8 $\mu$m emission in a ring-like morphology on the Galactic plane ($l=10^\circ$--$65^\circ$ and $295^\circ$--$350^\circ$) and termed these objects as {\it Spitzer} bubbles. 
The 8 $\mu$m emission [from hot dust and polycyclic aromatic hydrocarbon (PAH) molecule] of {\it Spitzer} bubbles surrounds the radio continuum (from ionized gas) and the 24 $\mu$m emission (from warm dust). 
According to \citet{1977ApJ...214..725E}, ultraviolet radiation from massive stars creates an expanding H{\sc ii} region and the interstellar gas and dust can be collected in the circumference of the H{\sc ii} region. 
As a result, dense ring- or shell-like gas/dust clouds are formed, which collapse to form stars. 
This is called as the ``collect and collapse'' process. 
This process has been suggested by some observational studies to operate in {\it Spitzer} bubbles (e.g., \cite{2009A&A...496..177D, 2010A&A...523A...6D, 2010A&A...518L..81Z}). 

\begin{figure}[h]
 \begin{center}
   \plotone{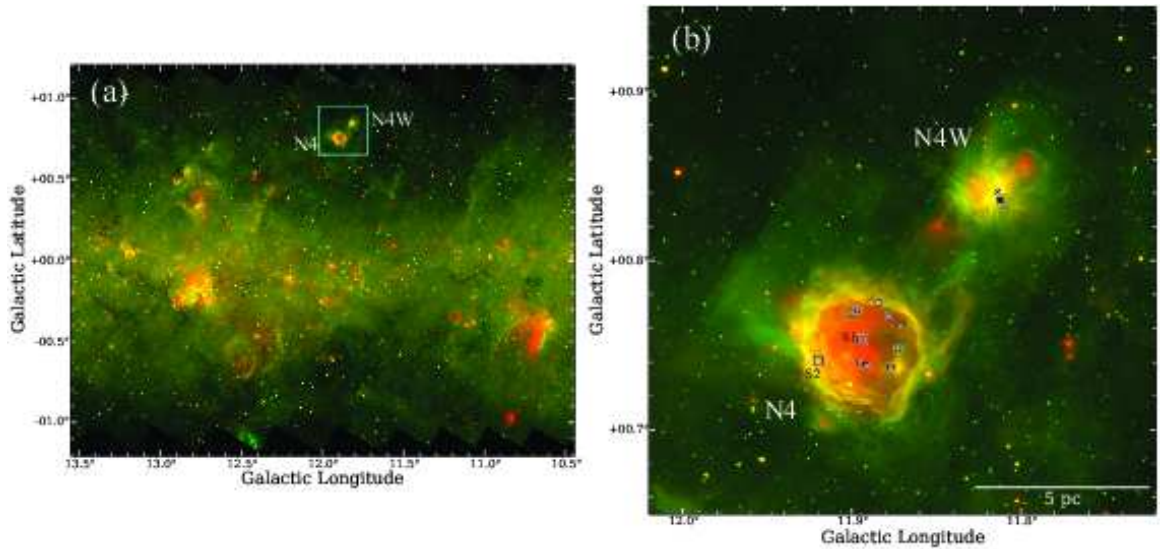}
 \end{center}
  \caption{(a) A composite color image of the {\it Spitzer}/MIPSGAL 24 $\mu$m (red) and {\it Spitzer}/GLIMPSE 8$\mu$m (green) emissions.  (b) Same as (a), but a closeup figure toward N4 and N4W. Square symbols represent the massive star candidates identified by \citet{2013RAA....13..921L}. In particular, the larger square symbols represent S1 and S2. Crosses represent the YSOs identified by \citet{2016ApJ...822..114C}.}\label{fig:RGB_all}
\end{figure}

Figure\,\ref{fig:RGB_all}(a) shows a composite color image of the {\it Spitzer}/MIPSGAL 24 $\mu$m (red) and {\it Spitzer}/GLIMPSE 8$\mu$m (green) emissions around N4 and N4W. 
N4 and N4W are located a little far from the middle of the Galactic plane, and seem to be relatively isolated objects from the surrounding H{\sc ii} regions. 
Figure\,\ref{fig:RGB_all}(b) shows a closeup figure of Figure\,\ref{fig:RGB_all}(a). 
N4 has an almost complete ring-like structure in the 8 $\mu$m emission and the H{\sc ii} region inside the ring.
In previous studies, molecular clouds associated with N4 have been detected and the clouds show a ring-like structure (\cite{2013RAA....13..921L}), and their radial velocities ($V_{\rm LSR}$) are centered on $\sim$25\,km\,s$^{-1}$.
According to the parallax-based distance estimator (\cite{2016ApJ...823...77R}), this $V_{\rm LSR}$ and direction corresponds to a probable distance of 2.80\,kpc$\pm$0.30\,kpc, and hence, we adopted a distance of 2.8\,kpc to N4. 
The radius of the ring is $\sim$2$'$, which corresponds to $\sim$1.6\,pc. 
The estimated total Lyman continuum [log($N_{\rm Lym}$\,s$^{-1}$) = 48.18] indicates that a main O8.5--O9V star is responsible for the ionization of N4 (\cite{2016ApJ...818...95L}).

\citet{2010ApJ...716.1478W} investigated the distribution of young stellar object (YSO) candidates around N4, and found that there does not appear to be an overdensity of YSOs along the shell. 
Therefore \citet{2010ApJ...716.1478W} suggested that there is no evidence for the triggered star formation via the ``collect and collapse'' process, although they claimed that the triggered star formation could not be ruled out because YSO samples are not complete in this region. 
On the contrary, by observing the $^{12}$CO ($J$=1--0), $^{13}$CO ($J$=1--0), and C$^{18}$O ($J$=1--0) emissions, \citet{2013RAA....13..921L} found an expanding motion of the molecular clouds associated with N4. 
They suggested that the formation of a massive star candidate (labeled as S2 in \citet{2013RAA....13..921L}) on the ring may be triggered by the expansion of the H{\sc ii} region formed by a massive star candidate (labeled as S1 in \citet{2013RAA....13..921L}); however, the formation mechanism of S1, which is the star exciting N4 and located inside the ring, remains unclear in this scenario.

N4W, located $\sim 5'$ west of N4, is a star-forming clump hosting YSOs and a submillimeter source (\cite{2016ApJ...822..114C}). 
\citet{2016ApJ...822..114C} identified four YSOs with at least intermediate mass in the innermost area by the observations of J, H, and Ks bands and found that these YSOs are almost coeval. 

\subsection{Paper overview}
In this paper, we report an observational study of the {\it Spitzer} bubble N4 and N4W using the $^{12}$CO ($J$=1--0) and $^{13}$CO ($J$=1--0) dataset obtained via the FUGIN project (\cite{2016SPIE.9914E..1ZM, 2017PASJ...69...78U}), whose spatial resolution is approximately 3\,times higher than the previous CO observations, to investigate the formation process of N4 and N4W. 
Section\,\ref{sec:Dat} describes these datasets. 
In Section\,\ref{sec:Res}, we describe the large-scale CO distribution (Subsection\,\ref{sec:COdist}), present the velocity structure of the molecular clouds (Subsection\,\ref{sec:COvelo}), compare the $^{12}$CO ($J$=1--0) emission with the $^{12}$CO ($J$=3--2) archive data obtained using the James Clerk Maxwell Telescope (JCMT; Subsection\,\ref{sec:COratio}), and estimate the physical parameters of the molecular outflow associated with N4W (Subsection\,\ref{sec:outflow}). 
In Section\,\ref{sec:Dis}, we discuss massive star formation and the H{\sc ii} region in N4 via a comparison with other massive-star forming regions, and the star formation in N4W.

\section{Datasets}\label{sec:Dat}

Observations of N4 were conducted as a part of the FUGIN project (\cite{2017PASJ...69...78U}) \footnote{http://jvo-dev.mtk.nao.ac.jp/portal/nobeyama/fugin.do} using the Nobeyama Radio Observatory (NRO) 45-m telescope.
Details of the observations, calibration, and data reduction are summarized in \citet{2017PASJ...69...78U}. 
In this study, we used the data of the $^{12}$CO ($J$=1--0) and $^{13}$CO ($J$=1--0) emissions covering $l = 12\fdg 05$--$11\fdg 70$ and $b = 0\fdg 65$--$1\fdg 00$ ($0\fdg 35 \times 0\fdg 35$).
The beam size of the NRO 45-m telescope is $\sim 15''$ at 115\,GHz, and the effective angular resolution is $\sim 20''$ \textcolor{black}{, which is due to the scanning pattern of the observations (\cite{2017PASJ...69...78U})}. 
The Spectral Analysis Machine for the 45-m telescope (SAM45) spectrometer (\cite{2011KUNO}) was used with a frequency resolution of 244.14\,kHz, and the effective velocity resolution was 1.3\,km\,s$^{-1}$ at 115\,GHz.
The typical system noise temperatures ($T_{\rm sys}$), including atmosphere, were $\sim$150\,K and $\sim$250\,K at 110 and 115\,GHz, respectively.  
\textcolor{black}{$T_{\rm sys}$ at 115\,GHz is higher than that of 110\,GHz because of effects of atmosphere.}
The final cube data comprise spatial grids of $8.5'' \times 8.5''$ and velocity channels of 0.65\,km\,s$^{-1}$.
The final root-mean-square noise temperature $T_{\rm{rms}}$ in $T_{\rm{mb}}$ scale are 1.5\,K and 0.7\,K per velocity channel for the $^{12}$CO ($J$=1--0) and $^{13}$CO ($J$=1--0), respectively, after the intensity calibration. 
Based on the observations of standard sources, the intensity variations were less than 10\%--20\% and 10\% for $^{12}$CO ($J$=1--0) and $^{13}$CO ($J$=1--0), respectively. 

We used the $^{12}$CO ($J$=3--2) archive data obtained with the Heterodyne Array Receiver Programme (HARP) installed on the JCMT.
The observations covered a $0\fdg 15 \times 0\fdg 15$ area, including N4, but not including N4W.
The data have an angular resolution of 14$''$ and a velocity resolution of 0.44 km s$^{-1}$.
At 345\,GHz, the main beam efficiency for HARP is 0.64$\pm$0.10. 

To improve the signal-to-noise ratio and compare the $J$=1--0 data with the $J$=3--2 data at the same angular resolution, we convolved the dataset using a Gaussian function to be FWHM  30$''$ for both the $J$=1--0 data data and the $J$=3--2 data. 
We also convolved the dataset for the velocity axis to be at a resolution of 1.3\,km\,s$^{-1}$ using the same method.


\section{Results}\label{sec:Res}

\subsection{Large-scale CO distribution}\label{sec:COdist}
\begin{figure}[h]
 \begin{center}
   \plotone{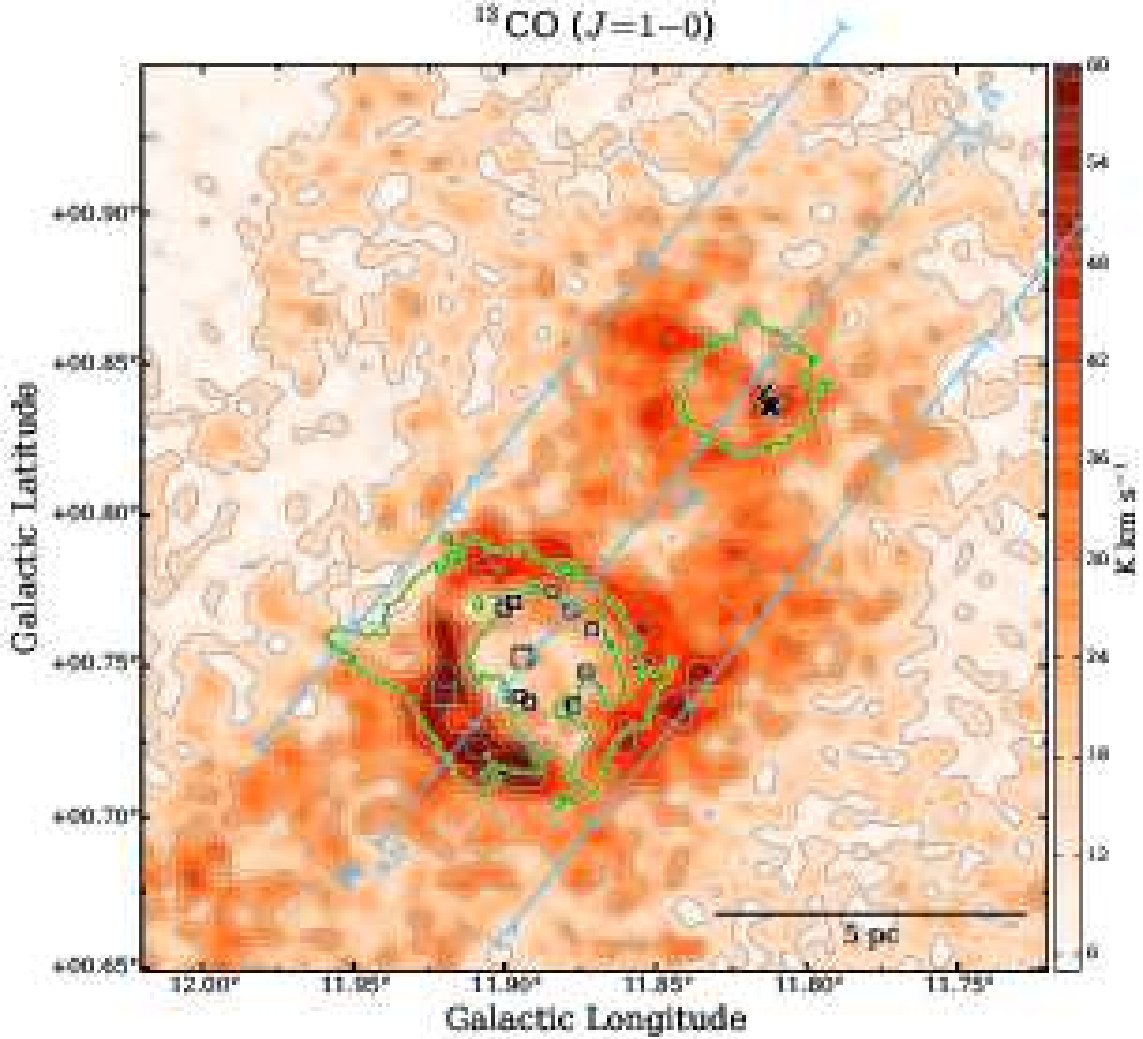}
 \end{center}
  \caption{Integrated intensity map of the $^{13}$CO ($J$=1--0) emission between the velocities of 12 and 37\,km\,s$^{-1}$. Black contours are plotted at every 12\,K\,km\,s$^{-1}$ from 12\,K\,km\,s$^{-1}$ ($\sim 5\sigma$). Green contours indicate the {\it Spitzer}/GLIMPSE 8$\mu$m intensity at 100 \,MJy\,str$^{-1}$. Blue lines indicate the direction and the width of the $p$--$v$ diagram shown in Figure\,\ref{fig:pvs}. Square symbols represent the massive star candidates identified by \citet{2013RAA....13..921L}.}\label{fig:integ_all}
\end{figure}

Figure\,\ref{fig:integ_all} shows the integrated intensity map of the $^{13}$CO ($J$=1--0) emission between the velocities of 12 and 37\,km\,s$^{-1}$, which covers the entire velocity range of N4 and N4W. 
The distribution of CO gas is consistent with previous studies (\cite{2013RAA....13..921L}) with a resolution of $>52''$.
The triangular symbols represent the YSOs identified by \citet{2016ApJ...818...95L} in N4. 
Most YSOs were distributed near the ring. 
Massive star candidates inside the ring were identified using near-infrared and mid-infrared data in N4 (\cite{2013RAA....13..921L}) and represented by square symbols. 
In N4, the molecular clouds seem to be distributed along the infrared 8 $\mu$m ring shown in Figure \ref{fig:RGB_all}, suggesting an association between the distributions of molecular clouds and dust. 
Black crosses represent intermediate-mass YSOs (\cite{2016ApJ...822..114C}) in N4W. 
Molecular clouds in N4W are associated with the infrared emission (8 $\mu$m and 24 $\mu$m) and the YSOs. 

The maximum column densities ($N({\rm H_2})$) were estimated to be $1.1 \pm 0.2 \times 10^{23}$\,cm$^{-2}$ and $0.8 \pm 0.1 \times 10^{23}$\,cm$^{-2}$ at 12 -- 37\,km\,s$^{-1}$ for N4 and N4W, respectively, which were derived from the $^{13}$CO ($J$=1--0) intensity and the local thermodynamic equilibrium (LTE) analysis. 
The estimated errors are due to mainly the calibration errors in the CO dataset (the same hereinafter).
In this derivation, we assumed that the $^{12}$CO ($J$=1--0) emission lines are optically thick and that the excitation temperatures ($T_{\rm ex}$) were derived from the $^{12}$CO ($J$=1--0) peak brightness temperatures for each pixel (the derived $T_{\rm ex}$ is typically 10--40\,K). 
We adopted an abundance ratio of [$^{12}$CO]/[$^{13}$CO]$=77$ (\cite{1994ARA&A..32..191W}) and a fractional $^{12}$CO abundance of $X$($^{12}$CO) = [$^{12}$CO]/[H$_2$]$=10^{-4}$ (\cite{1982ApJ...262..590F, 1984ApJS...56..231L}), and thus, $X$($^{13}$CO) = [$^{13}$CO]/[H$_2$]$=1.3\times10^{-6}$. 
Molecular masses were estimated to be $2.8 \pm 0.4 \times 10^{4}$\,$M_{\odot}$ and $2.1 \pm 0.3 \times 10^{4}$\,$M_{\odot}$ at 12 -- 37\,km\,s$^{-1}$ for N4 and N4W, respectively. 

\begin{figure}[h]
 \begin{center}
   \includegraphics[width=18cm]{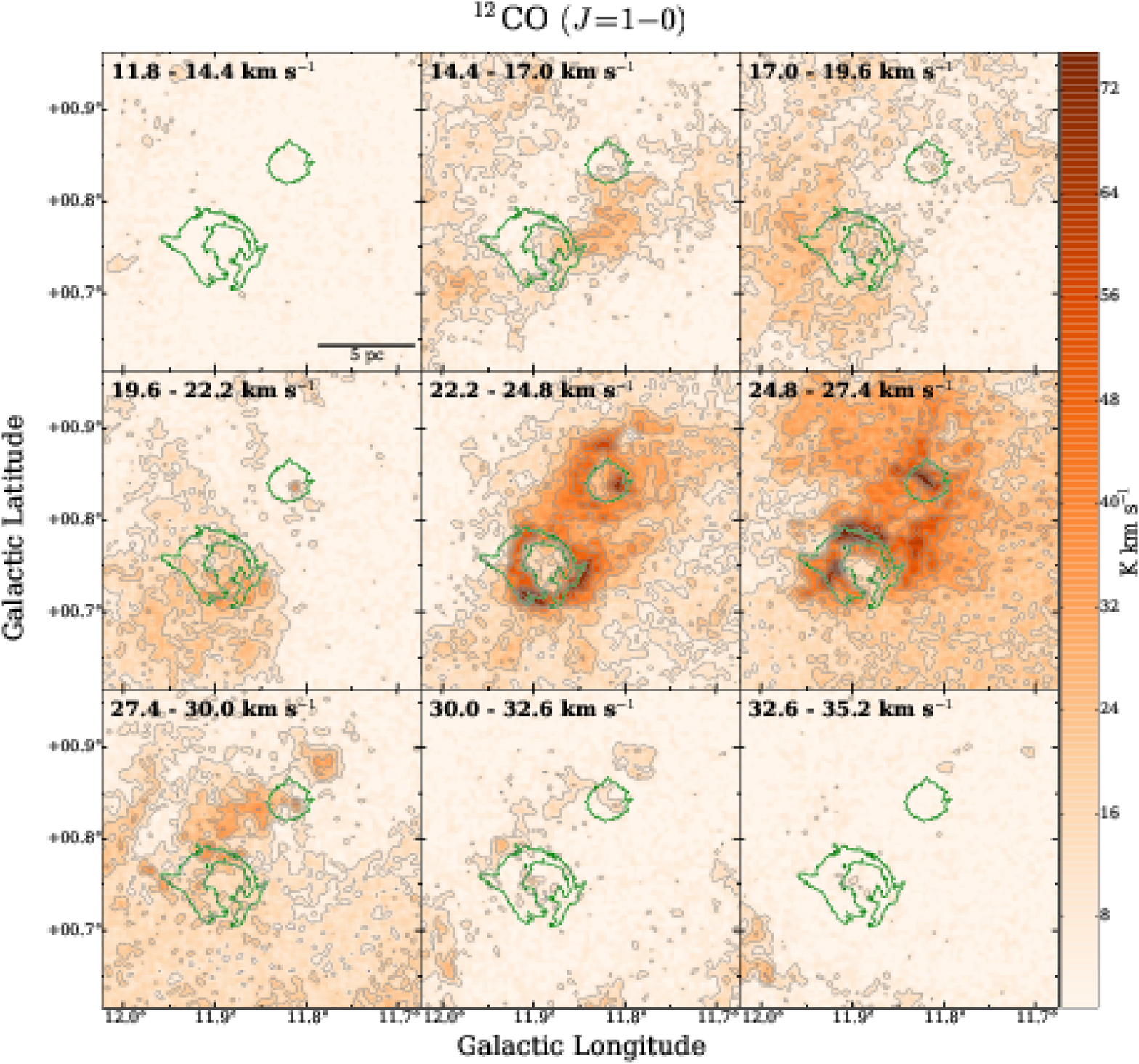}
 \end{center}
  \caption{The velocity channel maps of the $^{12}$CO ($J$=1--0) emissions. Integration range of each panel is presented in the top-left corner of the panel. Black contours are drawn at every 8.5\,K\,km\,s$^{-1}$ from 8.5\,K\,km\,s$^{-1}$ ($\sim 5\sigma$). Square symbols represent the massive star candidates identified by \citet{2013RAA....13..921L}. Green contours indicate the {\it Spitzer}/GLIMPSE 8$\mu$m intensity at 100\,MJy\,str$^{-1}$. }\label{fig:12COchmap}
\end{figure}

Figure\,\ref{fig:12COchmap} shows the large-scale velocity channel maps of the $^{12}$CO ($J$=1--0) emissions at a velocity step size of 2.6\,km\,s$^{-1}$. 
We also present the velocity channel maps of the $^{13}$CO ($J$=1--0) emissions in Figure\,\ref{fig:13COchmap} in the appendix for supplements.
The $^{12}$CO ($J$=1--0) emissions show extended gas distributions, whereas the $^{13}$CO ($J$=1--0) emissions show some clumpy structure. 
In the 22.2 -- 24.8\,km\,s$^{-1}$ panel, molecular ring-like structures were seen not only in N4 but also in N4W. 
In addition to the 22.2 -- 27.4\,km\,s$^{-1}$ panel, relatively diffuse components were observed in the 14.4 -- 17.0\,km\,s$^{-1}$, 17.0 -- 22.2\,km\,s$^{-1}$, and 27.4 -- 30.0\,km\,s$^{-1}$ panels.

\subsection{Velocity structure of the molecular clouds}\label{sec:COvelo}
\begin{figure}[h]
 \begin{center}
   \plotone{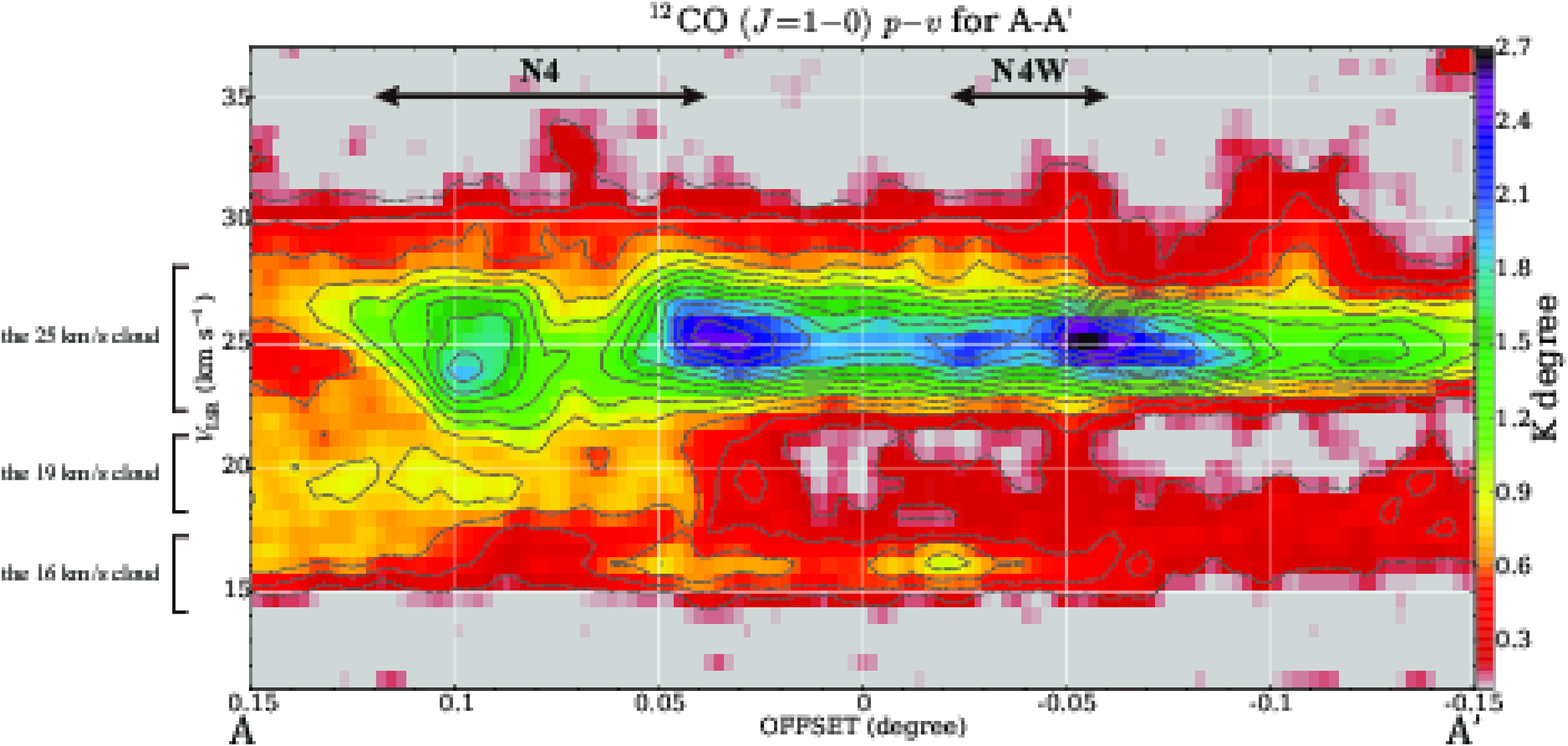}
 \end{center}
  \caption{Position--Velocity ($p-v$) diagram of the $^{12}$CO ($J$=1--0) emissions along the blue line shown in Figure\,\ref{fig:integ_all} from A to A$'$ with an integrated width of 6$'$ (between the upper and lower blue lines). Contours are plotted at every 0.1\,K\,degree from 0.1\,K\,degree ($\sim 5\sigma$). The black arrows indicate the area Spitzer/GLIMPSE 8μm intensity at $>$\,$\sim$\,100 M\,Jy\,str$^{-1}$ shown in Figure\,\ref{fig:integ_all}. }\label{fig:pvs}
\end{figure}

Figure\,\ref{fig:pvs} shows the position-velocity ($p$--$v$) diagram for the $^{12}$CO ($J$=1--0) emissions along the blue line in Figure\,\ref{fig:integ_all} with a width of 6$'$. 
In Figures\,\ref{fig:12COchmap} and \ref{fig:pvs}, we identified three discrete intensity peaks with velocities of $\sim$16, 19, and 25\,km\,s$^{-1}$ (hereinafter termed as ``the 16-km\,s$^{-1}$ cloud'', ``the 19-km\,s$^{-1}$ cloud'', and ``the 25-km\,s$^{-1}$ cloud'', respectively). 
In Figure\,\ref{fig:12COchmap}, these three components are shown in the 14.4--17.0, 17.0--22.2, and 22.2--27.4 \,km\,s$^{-1}$ panels, respectively.
The 16-km\,s$^{-1}$ cloud and the 19-km\,s$^{-1}$ cloud have not been recognized in previous observations of molecular lines. 
Furthermore, we found that the 16-km\,s$^{-1}$ cloud and the 25-km\,s$^{-1}$ cloud are connected \textcolor{black}{significantly ($>5\,\sigma$)} at an OFFSET value of $\sim 0 \fdg 05$ and an OFFSET value of $\sim -0 \fdg 05$ in Figure\,\ref{fig:pvs}. 
These clouds are therefore probably interacting with each other.
In addition, a diffuse component with a velocity of $\sim$\,29\,km\,s$^{-1}$ (i.e., the 29-km\,s$^{-1}$ cloud) can be observed in Figure\,\ref{fig:12COchmap} \textcolor{black}{and Figure\,\ref{fig:13COchmap} in Appendix}, although this component is blended with the 25-km\,s$^{-1}$ cloud in the $p$--$v$ diagram.


Figure\,\ref{fig:integ_ratio}(a) shows the $^{12}$CO ($J$=1--0) integrated intensity of the 16-km\,s$^{-1}$ cloud with an integrated velocity range from 14.4 to 17.0\,km\,s$^{-1}$.
The size of the 16-km\,s$^{-1}$ cloud is $\sim$5\,pc\,$\times$3\,pc. 
The left end of the cloud in the map is overlapped with the center of the infrared 8 $\mu$m ring, where the massive star candidates are identified. 
The mass and maximum column density ($N({\rm H_2})$) of the 16-km\,s$^{-1}$ cloud are estimated to be $0.12 \pm 0.02 \times 10^{4}$\,$M_{\odot}$ and $0.7\pm 0.1 \times 10^{22}$\,cm$^{-2}$, respectively. 

Figure\,\ref{fig:integ_ratio}(b) shows the $^{12}$CO ($J$=1--0) integrated intensity of the 19-km\,s$^{-1}$ cloud with an integrated velocity range from 17.7 to 20.2\,km\,s$^{-1}$.
This cloud extends around the infrared 8 $\mu$m ring, and the size of the cloud is approximately 8\,pc\,$\times$8\,pc. 
The mass and maximum $N({\rm H_2})$ of the 19-km\,s$^{-1}$ cloud are estimated to be $0.43 \pm 0.06 \times 10^{4}$\,$M_{\odot}$ and $0.8 \pm 0.1 \times 10^{22}$\,cm$^{-2}$, respectively. 

Figure\,\ref{fig:integ_ratio}(c) shows the $^{12}$CO ($J$=1--0) integrated intensity of the 25-km\,s$^{-1}$ cloud within an integrated velocity range between 22.2 and 27.4\,km\,s$^{-1}$.
This cloud also has a distribution to likely trace the ring structure in N4. 
The mass and maximum $N({\rm H_2})$ of the 25-km\,s$^{-1}$ cloud are estimated to be $3.8 \pm 0.6 \times 10^{4}$\,$M_{\odot}$ and $1.0 \pm 0.2\times 10^{23}$\,cm$^{-2}$, respectively.

\subsection{$^{12}$CO ($J$=3--2)/($J$=1--0) intensity ratio}\label{sec:COratio}
\begin{figure}[h]
 \begin{center}
   \plotone{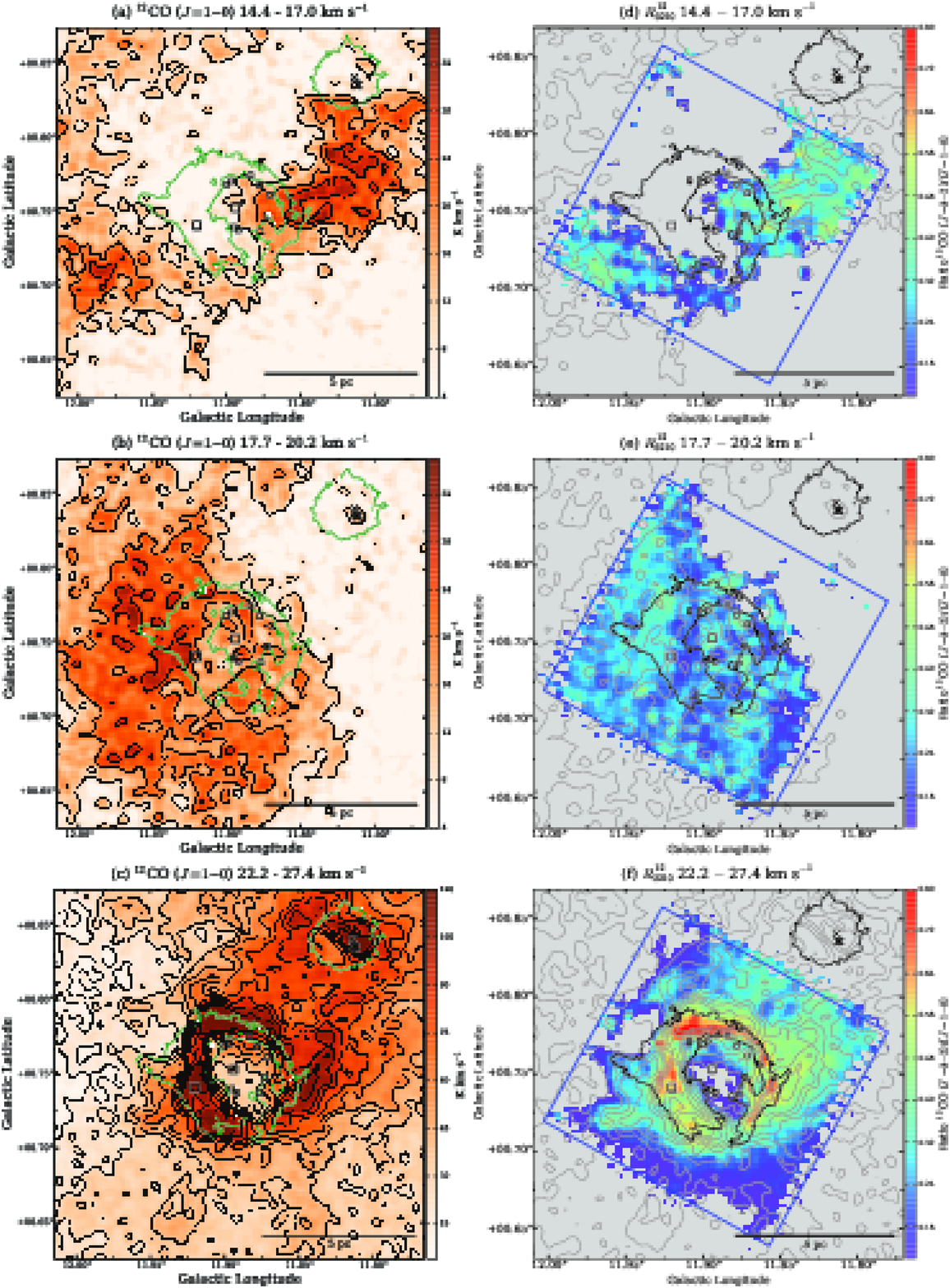}
 \end{center}
 \caption{(a)--(c) Integrated intensity of the $^{12}$CO ($J$=1--0) of the 16-, 19-, and 25-km\,s$^{-1}$ clouds, respectively. Green contours indicate the {\it Spitzer}/GLIMPSE 8$\mu$m intensity at 100\,MJy\,str$^{-1}$. (d)--(f) Integrated intensity ratio $^{12}$CO ($J$=3--2)/$^{12}$CO ($J$=1--0) ($R^{12}_{3210}$) map of the 16-, 19-, and 25-km\,s$^{-1}$ clouds, respectively. Thin black contours indicate the $^{12}$CO ($J$=1--0) integrated intensity, and plotted every 8\,K\,degree ($\sim 5\sigma$) from 10\,K\,degree, every 8\,K\,degree ($\sim 5\sigma$) from 10\,K\,degree, and every 12\,K\,degree ($\sim 5\sigma$) from 12\,K\,degree, respectively. Blue large square indicates the area of the $^{12}$CO ($J$=3--2) data. Square symbols represent the massive star candidates identified by \citet{2013RAA....13..921L}. }\label{fig:integ_ratio}
\end{figure}

The color scale in Figures\,\ref{fig:integ_ratio}(d)--\ref{fig:integ_ratio}(f) shows the integrated intensity ratio $^{12}$CO ($J$=3--2)/$^{12}$CO ($J$=1--0) (hereinafter denoted by $R^{12}_{3210}$) of the 16-km\,s$^{-1}$ cloud, the 19-km\,s$^{-1}$ cloud, and the 25-km\,s$^{-1}$ cloud, respectively.
The intensity ratio between two different rotational transitions reflects the kinematic temperature and/or the density of the gas. 
In Figure\,\ref{fig:integ_ratio}, $R^{12}_{3210}$ is relatively high ($>$0.7) at the ring structure of the 25-km\,s$^{-1}$ cloud, probably suggesting that the gas has been heated due to radiation from massive stars within the ring \textcolor{black}{and/or the volume density of the gas is high}.
On the contrary, $R^{12}_{3210}$ of the 16-km\,s$^{-1}$ cloud and the 19-km\,s$^{-1}$ cloud were lower than that of the 25-km\,s$^{-1}$ cloud, typically $\sim$0.45 and $\sim$0.30, respectively.

\subsection{Molecular outflow in N4W}\label{sec:outflow}
\begin{figure}[h]
 \begin{center}
   \plotone{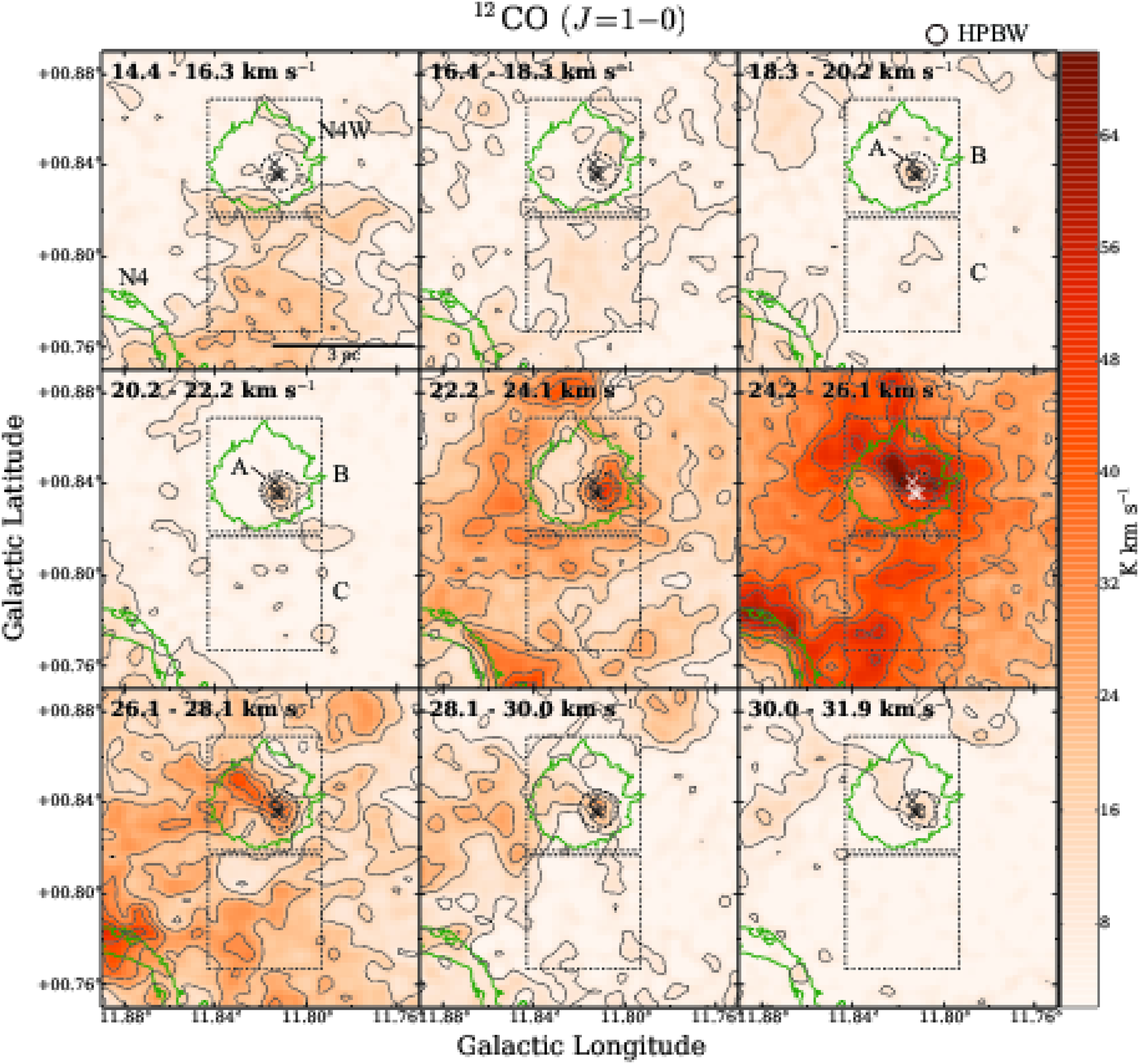}
 \end{center}
  \caption{Velocity channel maps of the $^{12}$CO ($J$=1--0) emissions toward N4W. Integration range of each panel is presented in the top-left corner of the panel. Crosses represent the YSO candidates identified by \citet{2016ApJ...822..114C}. Black contours are plotted at every 7.5\,K\,km\,s$^{-1}$ from 7.5\,K\,km\,s$^{-1}$ ($\sim 5\sigma$). Green contours indicate the {\it Spitzer}/GLIMPSE 8$\mu$m intensity at 100\,MJy\,str$^{-1}$. Dotted circle A shows the area we used in Figure\,\ref{fig:N4W_12CO_chmap_2}. Dotted squares B and C show the area we used in Figure\,\ref{fig:N4W_12CO_chmap_3}. }\label{fig:N4W_12CO_chmap_1}
\end{figure}

\begin{figure}[h]
 \begin{center}
   \includegraphics[width=9cm]{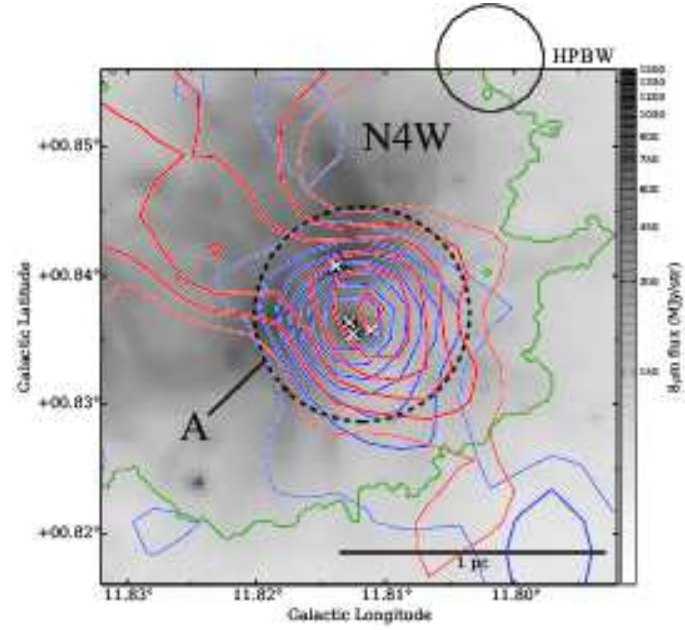}
 \end{center}
  \caption{Gray scale shows the intensity of the {\it Spitzer}/GLIMPSE 8$\mu$m (green) emissions toward N4W, and the green contours indicate 100\,MJy\,str$^{-1}$. Blue and red contours show the integrated intensity of the $^{12}$CO ($J$=1--0) emissions with the velocity range of 18.3 -- 22.2\,km\,s$^{-1}$ and  28.7 -- 32.6\,km\,s$^{-1}$, respectively. The contours are plotted at every 5\,K\,km\,s$^{-1}$ from 10\,K\,km\,s$^{-1}$ ($\sim 5\sigma$). Crosses represent the YSO candidates identified by \citet{2016ApJ...822..114C}. }\label{fig:N4W_outflow}
\end{figure}

Figure\,\ref{fig:N4W_12CO_chmap_1} shows the velocity channel map of the $^{12}$CO ($J$=1--0) emission around N4W.  
A compact component with a large line width can be observed at $(l,b)=(11\fdg812, 0\fdg837)$ with the velocity center at $\sim$25\,km\,s$^{-1}$.
In this position, four intermediate-mass YSOs, indicated by crosses, and their outflows have been identified by a previous study (\cite{2016ApJ...822..114C}).
The compact component with a broad line width located on the YSOs is considered to be a molecular outflow from one of the YSOs.
Figure\,\ref{fig:N4W_outflow} shows the distribution of the molecular outflow in two velocity ranges. 
The peak positions of the blue shifted and red shifted outflow lobes are slightly shifted from each other, although the separation is smaller than the resolution of the CO data.

\begin{figure}[h]
 \begin{center}
   \includegraphics[width=9cm]{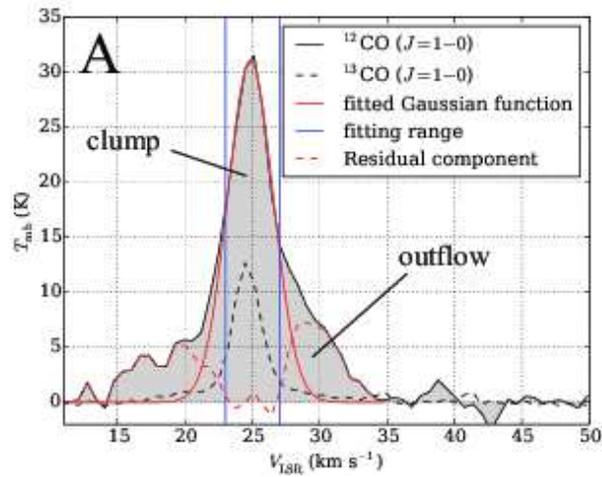}
 \end{center}
  \caption{The averaged spectra in the circle A in Figure\,\ref{fig:N4W_12CO_chmap_1}. }\label{fig:N4W_12CO_chmap_2}
\end{figure}

We derived the physical parameters of this molecular outflow, e.g., mass ($M_{\rm flow}$), kinetic energy ($E_{\rm flow}$), momentum ($P_{\rm flow}$), dynamical timescale ($t_{\rm dyn}$), and mechanical luminosity ($L_{\rm m}$), by employing the method of \citet{2017ApJ...840..111T}, which assumes the LTE condition. 
Because the CO spectra of the outflow are blended with respect to the molecular clump of the 25-km\,s$^{-1}$ cloud, we first determined the velocity range of the CO spectra of the clump at the outflow position. 
The spatial extension of the clump was identified by drawing a contour onto the $^{13}$CO ($J$=1--0) integrated intensity map (integrated from 12\,km\,s$^{-1}$ to 37\,km\,s$^{-1}$) at the two-third level of the peak intensity delineated by circle A in Figures\,\ref{fig:N4W_12CO_chmap_1} and \ref{fig:N4W_outflow}. 
Figure\,\ref{fig:N4W_12CO_chmap_2} show the averaged CO spectra in the circle A. 
A fit with a Gaussian function was performed for the $^{12}$CO ($J$=1--0) spectrum averaged over circle A, which provided a systemic velocity of 25\,km\,s$^{-1}$ and a velocity width of 4.0\,km\,s$^{-1}$. 
We estimated the mass of the molecular clump, which is possibly host the YSOs, of $4.1 \pm 0.6 \times 10^{2}\,M_{\odot}$ in the velocity range 22.2 -- 27.4\,km\,s$^{-1}$. 
This value is consistent with the estimation of \citet{2016ApJ...822..114C}. 
We regard the residuals of the CO spectra and the Gaussian function as the molecular outflow components in the CO spectra (Figure\,\ref{fig:N4W_12CO_chmap_2}). 
Since other components are blended in the spectra in the lower velocity side of the outflow, we derived the physical parameters only the higher velocity side of the outflow.
Although the 29\,km\,s$^{-1}$ component is blended into the higher velocity side with the outflow component, we ignore the 29\,km\,s$^{-1}$ component because its intensity is low. 
Assuming that the inclination of the line-of-sight direction is \textcolor{black}{0$^{\circ}$ or 45$^{\circ}$}, we estimated $M_{\rm flow}=\sum_{i} M_{{\rm flow, }i}= 0.5 \pm 0.1 \times10^2\,M_{\odot}$, \textcolor{black}{total $P_{\rm flow}=\sum_{i} M_{{\rm flow, }i}v_i=249\pm 75 \,M_{\odot}$\,km\,s$^{-1}$ for 0$^{\circ}$ and $352\pm 53 \,M_{\odot}$\,km\,s$^{-1}$ for 45$^{\circ}$, and total $E_{\rm flow}=\sum_{i} M_{{\rm flow, }i}v_i^2=0.9 \pm 0.1 \times10^{46}$\,erg for 0$^{\circ}$ and $1.8 \pm 0.3 \times10^{46}$\,erg for 45$^{\circ}$}, where $i$ refers to velocity channels. 
\textcolor{black}{According to a statistical study of YSOs in our Galaxy by \citet{2004A&A...426..503W}, the estimated $M_{\rm flow}$ is approximately the intermediate value between the low mass group and the high mass group in the study, if we assume that the outflow stems from a single YSO.}


\clearpage

\section{Discussion}\label{sec:Dis}

\subsection{Expanding motion of the molecular gas in N4}\label{sec:Dis_exp}
\citet{2013RAA....13..921L} observed N4 with $^{12}$CO ($J$=1--0), $^{13}$CO ($J$=1--0), and C$^{18}$O ($J$=1--0) and speculated that N4 is more likely an inclined expanding ring than a spherical-bubble, although they also noted that observations with higher resolution are necessary to confirm this speculation. 
\citet{2013RAA....13..921L} also speculated that the formation of S2 on the ring was triggered by the compression due to the expanding motion of the ring. 
On the contrary, \citet{2017ApJ...838...80C} observed a magnetic field derived from near-IR polarization of reddened diskless stars located behind N4. 
They found that the direction of the magnetic field is curved and parallel to the ring and suggested that the star formation on the ring triggered by expanding motions might not easily occur because the estimated magnetic field is strong enough ($\sim$120\,$\mu$G).

\begin{figure}[h]
 \begin{center}
   \plotone{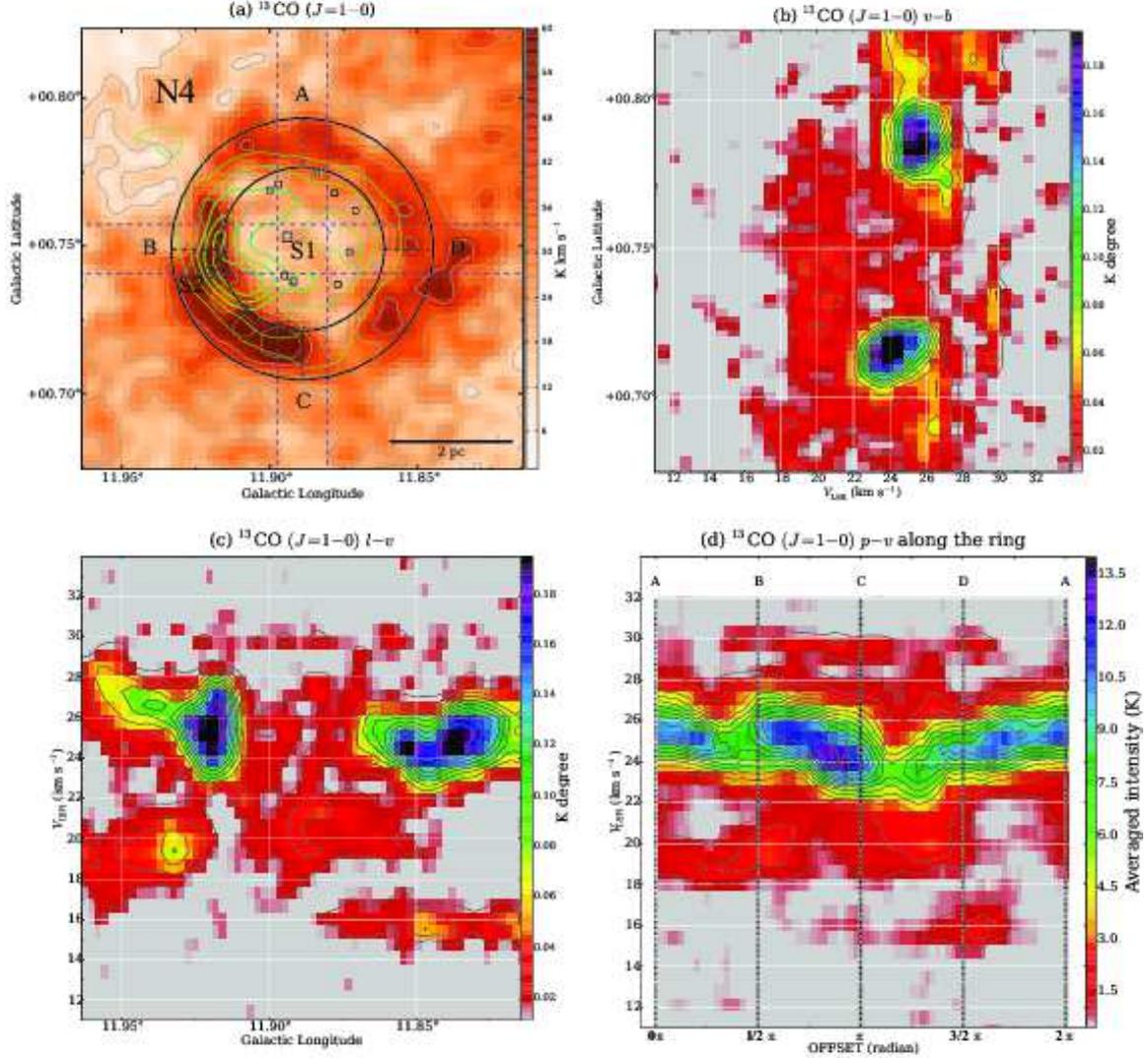}
 \end{center}
 \caption{(a) The $^{13}$CO ($J$=1--0) integrated intensity between the velocities of 12 and 37\,km\,s$^{-1}$. The green contours show the intensity of 20-cm radio continuum taken from the Multi-Array Galactic Plane Imaging Survey (MAGPIS, \cite{2006AJ....131.2525H}) archive, and are plotted by 0.75 ($\sim 5\sigma$), 1.95, 3.15, 4.35, and 5.55 mJy\,Beam$^{-1}$. Square symbols represent the massive star candidates identified by \citet{2013RAA....13..921L}. (b) The $v$--$b$ diagram of the $^{13}$CO ($J$=1--0) emission integrated between the vertical blue dotted-lines in Figure\,\ref{fig:circ_dist}(a). \textcolor{black}{Contours are plotted at every 0.018\,K\,degree from 0.018\,K\,degree ($\sim 5\sigma$).} (c) The $l$--$v$ diagram of the $^{13}$CO ($J$=1--0) emission integrated between the horizontal blue dotted-lines in Figure\,\ref{fig:circ_dist}(a). \textcolor{black}{Contours are plotted at every 0.018\,K\,degree from 0.018\,K\,degree ($\sim 5\sigma$).} (d) The $p$--$v$ diagram of the $^{13}$CO ($J$=1--0) emission along the ring with a width of 1$'$ indicated by the black circles in Figure \ref{fig:circ_dist}(a). \textcolor{black}{Contours are plotted at every 1.0\,K from 1.0\,K ($\sim 5\sigma$).}}\label{fig:circ_dist}
\end{figure}

We investigate the detailed velocity structure of the 25-km\,s$^{-1}$ cloud by using our approximately 3\,times high-angular resolution CO dataset. 
Figure \ref{fig:circ_dist}(a) shows the $^{13}$CO ($J$=1--0) integrated intensity between the velocities of 12 and 37\,km\,s$^{-1}$. 
Figures \ref{fig:circ_dist}(b) and \ref{fig:circ_dist}(c) show the $v$--$b$ diagram and $l$--$v$ diagram of the $^{13}$CO ($J$=1--0) emission, respectively, integrated between the blue dotted-lines in Figure\,\ref{fig:circ_dist}(a). 
\textcolor{black}{The lowest contour indicates $\sim 5\sigma$ level.}
If the molecular gas in N4 has an expanding spherical-bubble structure, elliptical shapes should be observed in Figures \ref{fig:circ_dist}(b) and \ref{fig:circ_dist}(c) as suggested by Figure\,5 in \citet{2011ApJ...742..105A}, which is a model of an expanding spherical-bubble inside a turbulent medium.
However, we can not observe clear ellipse in Figures \ref{fig:circ_dist}(b) and \ref{fig:circ_dist}(c), which is consistent with the result of \citet{2013RAA....13..921L}.

On the other hand, Figure \ref{fig:circ_dist}(d) shows the $p$--$v$ diagram of the $^{13}$CO ($J$=1--0) emission along the ring with a width of 1$'$ (between the black circles in Figure \ref{fig:circ_dist}(a)). 
\textcolor{black}{The lowest contour indicates $\sim 5\sigma$ level.}
If the molecular gas in N4 has an expanding ring structure as proposed by \citet{2013RAA....13..921L}, Figure \ref{fig:circ_dist}(\textcolor{black}{d}) should show a sinusoidal wave with a length of 2$\,\pi$ radian (one cycle) unless the expanding motion is perpendicular to the line-of-sight. 
However, we can not find a clear sinusoidal wave of the 25-km\,s$^{-1}$ cloud in Figure \ref{fig:circ_dist}(d), though some velocity gradients are observed. 
For these reasons, we concluded that the molecular gas in N4 may not be expanding.

\subsection{Cloud--cloud collisions in N4 and N4W as an alternative scenario}\label{sec:Dis_ccc}
\subsubsection{N4}

As shown in Figures\,\ref{fig:12COchmap}, \ref{fig:integ_ratio}(c), and \ref{fig:integ_ratio}(f), the 25-km\,s$^{-1}$ cloud is clearly associated with N4. 
The 16-km\,s$^{-1}$ cloud is also possibly associated with N4 because it has slightly elevated $R^{12}_{3210}$, as shown in Figure\,\ref{fig:integ_ratio}(d). 
Meanwhile, it is not certain whether the 19-km\,s$^{-1}$ cloud is interacting with the 16-km\,s$^{-1}$ cloud and the 25-km\,s$^{-1}$ cloud. 
$R^{12}_{3210}$ of the 19-km\,s$^{-1}$ cloud is lower than those of the 16-km\,s$^{-1}$ cloud and the 25-km\,s$^{-1}$ cloud, and hence, it seems that the 19-km\,s$^{-1}$ cloud is not interacting with the H{\sc ii} region. 
The location of N4 is the inner Galaxy, $l=11\fdg8$, where heavy contamination is expected in these velocities.
Therefore, the 19-km\,s$^{-1}$ cloud is possibly not directly interacting with N4, but it is overlapping with other clouds at the line-of-sight. 

Herein, we found that the 16-km\,s$^{-1}$ cloud and the 25-km\,s$^{-1}$ cloud are connected in the $p$--$v$ diagram (Figure\,\ref{fig:pvs}) toward N4. 
This is a bridge feature, which is discussed as a possible observational signature of CCC by previous studies (e.g., \cite{2015MNRAS.450...10H, 2015MNRAS.454.1634H}). 
Figure\,\ref{fig:integ_overlay}(a) shows an integrated intensity map of the $^{12}$CO ($J$=1--0) emissions of the 25-km\,s$^{-1}$ cloud (color scale) and the 16-km\,s$^{-1}$ cloud (blue contours). 
The Galactic east end of the 16-km\,s$^{-1}$ cloud in the map is located at the center of the ring structure of the 25-km\,s$^{-1}$ cloud, where the massive star candidates have been identified. 
Figure\,\ref{fig:integ_overlay}(b) shows an integrated intensity map of the $^{13}$CO ($J$=1--0) emissions. 
At the Galactic west side of the 25-km\,s$^{-1}$ cloud in the map, the two clouds show spatially complementary distributions, which is also discussed as a possible observational signature of CCC (\cite{2018ApJ...859..166F}).  

Figure\,\ref{fig:pv_N4_N4W_1}(a) shows the $p$--$v$ diagram between B and B$'$ in Figure\,\ref{fig:integ_overlay}(a) through the 16-km\,s$^{-1}$ cloud and the 25-km\,s$^{-1}$ cloud.
A bridge feature connecting the 16-km\,s$^{-1}$ cloud and the 25-km\,s$^{-1}$ cloud can be observed, and it further shows a V-shaped structure (black dashed lines). 
Such V-shaped structures in $p$--$v$ diagrams have been observed in other CCC objects (e.g., \cite{2018ApJ...859..166F, 2018PASJ...70S..47O}). 
We can also observe the 19-km\,s$^{-1}$ cloud in the $p$--$v$ diagram, but this can be determined to be distinct from the bridge feature. 

We here test dynamical binding of the 16- and 25-km\,s$^{-1}$ clouds. 
If we tentatively assume that the two clouds are separated by 6\,pc (same as lengths of the 16- and 25-km\,s$^{-1}$ clouds in N4) in space and by $9\,\times \,\sqrt{2}$ km\,s$^{-1}$ in velocity (we also assume the viewing angle of the relative motion between the two clouds as 45$^{\circ}$ to the line-of-sight), the total mass required to gravitationally bind these two clouds can be calculated as $M=\frac{rv^2}{2G}=1.1\,\times 10^5M_{\odot}$. 
This is larger than the total molecular mass of N4 ($2.8\,\times 10^4M_{\odot}$) estimated in Section \ref{sec:COdist}, indicating that the co-existence of the two velocity clouds in N4 can not be interpreted as the gravitationally bound system.

For these reasons, we proposed a CCC between the 16-km\,s$^{-1}$ cloud and the 25-km\,s$^{-1}$ cloud in N4. 
After the collision, a cavity was created in the molecular clouds, which permitted the formation of one or more massive stars at the compressed layer. 
At present, gas near the collision interface of the 16-km\,s$^{-1}$ cloud is \textcolor{black}{broken up} and ionized via UV radiation from the massive star(s).


The timescale of the collision (between the time when collision occurred and the present time) in N4 can be approximated estimated from the size of the cavity and the relative velocity between the two clouds. 
The size of the cavity in the $l$--$b$ plane is $\sim$3\,pc. 
If we assume that the relative velocity parallel to the $l$--$b$ plane is same as the relative radial velocity and that the cavity is spherical, the estimated timescale of the collision is $3\,$pc$/9\,$km\,s$^{-1}=0.2$--$0.3$\,Myrs. 

The red and blue triangles in Figure\,\ref{fig:integ_overlay}(a) represent Class I YSOs and Class II YSOs, respectively, and the uncolored triangles represents transitional disk YSOs (\cite{2016ApJ...818...95L}). 
Class I and Class II YSOs are located at the extension line of the 16-km\,s$^{-1}$ cloud elongation and at the left edge of the 16-km\,s$^{-1}$ cloud in the map. 
The formation of these YSOs is \textcolor{black}{most likely} triggered by the collision of the 16-km\,s$^{-1}$ cloud and the 25-km\,s$^{-1}$ cloud, although the formation of the other YSOs in N4 could not has been triggered because the age of the transitional disk YSOs is generally greater than 1\,Myr (\cite{2011ApJ...732...24C}).

\subsubsection{N4W}
\begin{figure}[h]
 \begin{center}
   \includegraphics[width=18cm]{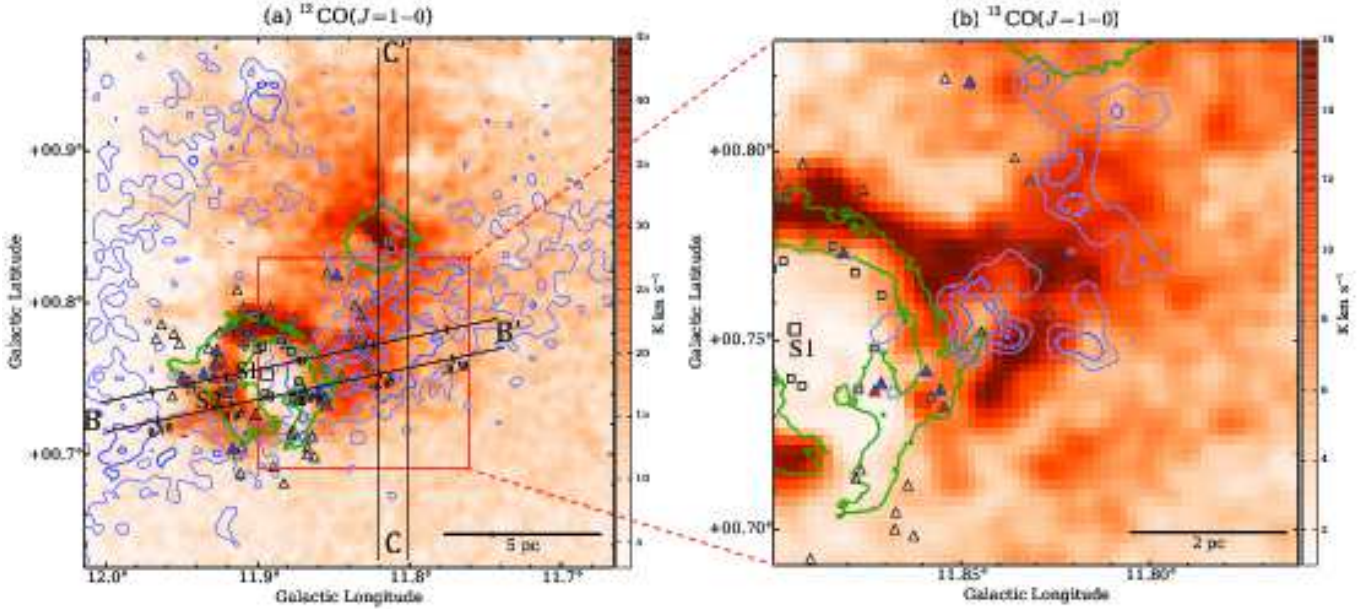}
 \end{center}
  \caption{(a) Integrated intensity map of the $^{12}$CO ($J$=1--0) emission with a velocity range from 24.8 to 26.1\,km\,s$^{-1}$ (color scale) (the 25-km\,s$^{-1}$ cloud) and from 15.1 to 16.4\,km\,s$^{-1}$ (blue contours) (the 16-km\,s$^{-1}$ cloud). Contours are plotted every 6\,K\,km\,s$^{-1}$ from 6\,K\,km\,s$^{-1}$ ($\sim 5\sigma$). Square symbols represent the massive star candidates identified by \citet{2013RAA....13..921L}. Triangle symbols represent the YSOs identified by \citet{2016ApJ...818...95L}, and red, blue, and non-colored indicate Class I, Class II, and transitional disk, respectively. Green contours show the {\it Spitzer}/GLIMPSE 8$\mu$m intensity at 100\,MJy\,str$^{-1}$  (b) Same as (a) but for the $^{13}$CO ($J$=1--0) emission \textcolor{black}{at the red square region in (a)}. Contours are plotted every 1\,K\,km\,s$^{-1}$ from 3\,K\,km\,s$^{-1}$ ($\sim 5\sigma$). }\label{fig:integ_overlay}
\end{figure}

\begin{figure}[h]
 \begin{center}
   \includegraphics[width=9cm]{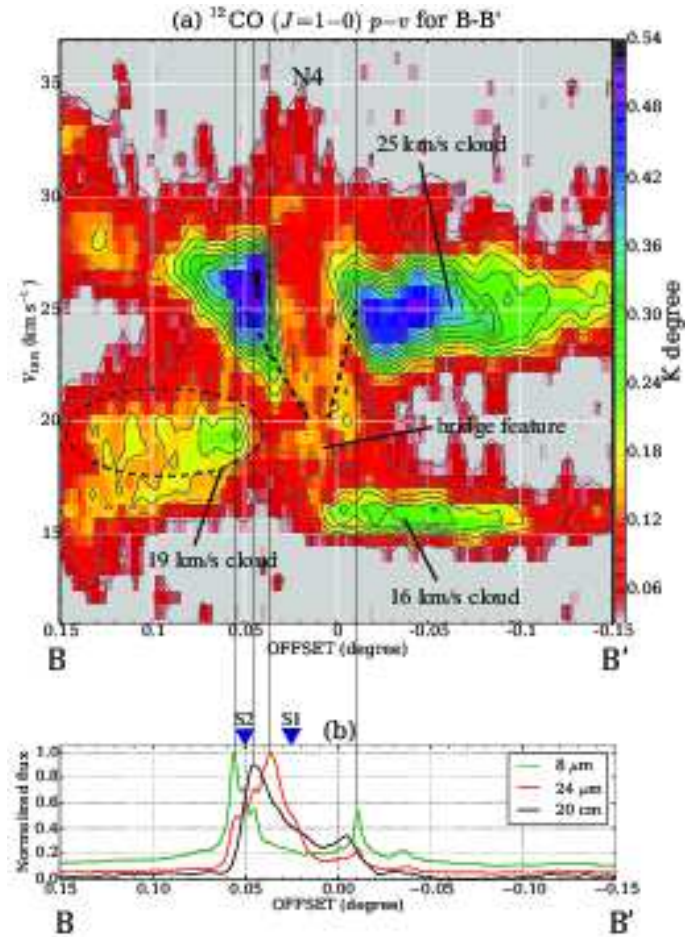}
 \end{center}
  \caption{(a) $p-v$ diagram of the $^{12}$CO ($J$=1--0) along black axis in Figure\,\ref{fig:integ_overlay}(a) from B to B$'$. Contours are drawn at every 0.08\,K\,degree from 0.08\,K\,degree ($\sim 10\sigma$). (b) Median intensity (perpendicular to the B--B$'$ line) of the {\it Spitzer}/MIPSGAL 24 $\mu$m (red) and {\it Spitzer}/GLIMPSE 8$\mu$m (green) along the B--B$'$ line in (a) with a width of 1.2$'$. }\label{fig:pv_N4_N4W_1}
\end{figure}

\begin{figure}[h]
 \begin{center}
   \includegraphics[width=9cm]{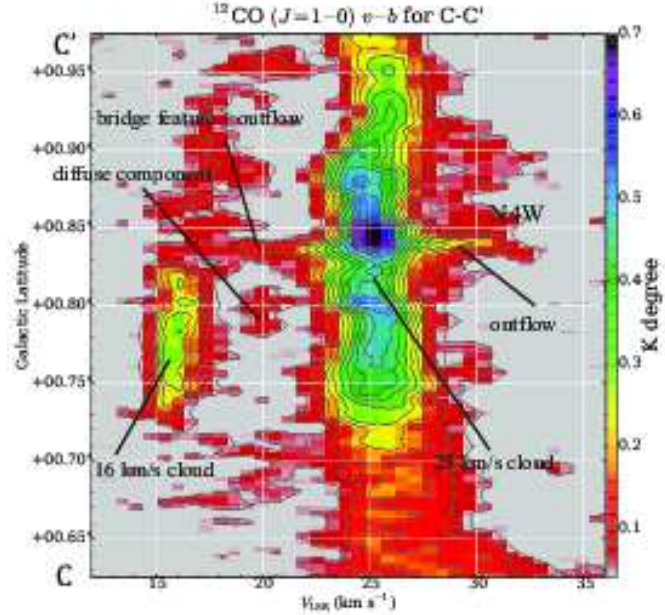}
 \end{center}
  \caption{$v-b$ diagram of the $^{12}$CO ($J$=1--0) from C to C$'$ in Figure\,\ref{fig:integ_overlay}(a). Contours are plotted at every 0.05\,K\,degree from 0.05\,K\,degree ($\sim 5\sigma$). }\label{fig:pv_N4_N4W_2}
\end{figure}

\begin{figure}[h]
 \begin{center}
   \plotone{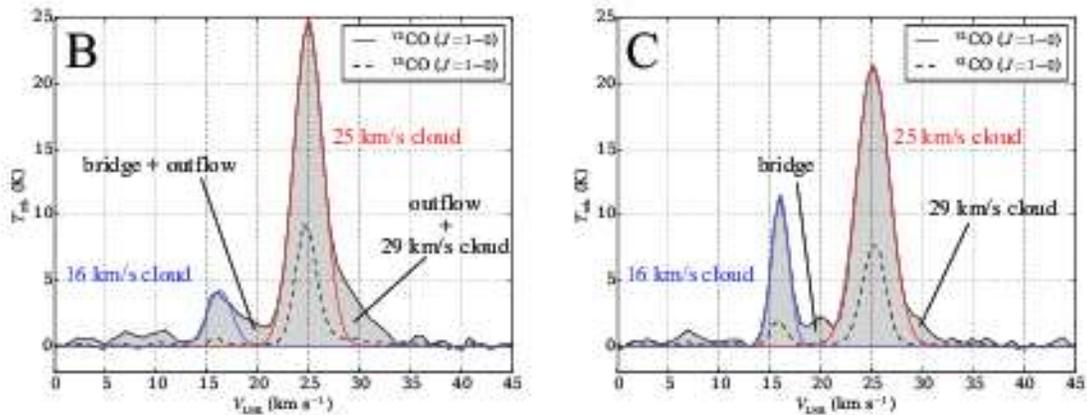}
 \end{center}
  \caption{Averaged spectra of $^{12}$CO ($J$=1--0) emission (solid line) and $^{12}$CO ($J$=1--0) emission (broken line) in the squares B (left) and C (right) in Figure\,\ref{fig:N4W_12CO_chmap_1}, respectively. }\label{fig:N4W_12CO_chmap_3}
\end{figure}

As seen in Figures\,\ref{fig:12COchmap}, and \ref{fig:integ_ratio}(c) the 25-km\,s$^{-1}$ cloud is associated not only with N4 but also with N4W. 
Figure\,\ref{fig:pv_N4_N4W_2} shows the $v$--$b$ diagram of the $^{12}$CO ($J$=1--0) emission from C to C$'$ in Figure\,\ref{fig:integ_overlay}(a).
We can see a diffuse component between the 16-km\,s$^{-1}$ cloud and the 25-km\,s$^{-1}$ cloud at a $b$ value of $\sim 0\fdg80$ \textcolor{black}{with the intensity of the $^{12}$CO ($J$=1--0) emission of $>5\sigma$}. 
Figure\,\ref{fig:N4W_12CO_chmap_3} shows the average spectra within the squares B and C in Figure\,\ref{fig:N4W_12CO_chmap_1}.
Although the high-velocity wing emission of the molecular outflows from the YSOs is confined to within the area of square B, a faint emission connecting the 16-km\,s$^{-1}$ cloud and the 25-km\,s$^{-1}$ cloud can be detected within the area of square C. 

For these reasons, the diffuse component between the two clouds may be a bridge feature similar to that observed in N4.
In other words, even in N4W, a similar scenario of massive star formation triggered by the collision of the 16-km\,s$^{-1}$ and the 25-km\,s$^{-1}$ clouds is conceivable.
Because the estimated $t_{\rm dyn}$ of the outflow from the YSOs is $<$\,0.1\,Myr (Subsection \ref{sec:outflow}) and H{\sc ii} regions in N4W have not grown compared to those of N4, the molecular clouds in N4W were collided probably after the collision in N4 in this scenario. 
\citet{2016ApJ...822..114C} suggested the possibility that the four YSOs in N4W are coeval.
The CCC scenario in N4W could explain the small age range of the reported YSOs since CCC can trigger star formation over a short time scale.
We speculate that the non-uniform cloud morphology and density caused massive star formations in two places (N4 and N4W) despite the single pair of collisions.

\subsection{Ages of the H{\sc ii} region in N4}\label{sec:Dis_age}
One can calculate a dynamical age of the H{\sc ii} region using the analytical model of the D-type expansion developed by \citet{1978ppim.book.....S}. 
The Lyman continuum photon flux of N4 was estimated to be $log(N_{\rm ly}\,{\rm s^{-1}})=48.18$ (\cite{2016ApJ...818...95L}). 
The initial volume density of the gas was estimated to be approximately 4$\times$10$^3$\,cm$^{-3}$ when assuming a uniform spherical distribution with a radius of 4\,pc with a total molecular mass of $\sim 2.8\times$10$^4$\,M$_{\odot}$.  
In addition, we assumed an electron temperature of 8000\,K. 
Given these parameters, the age of the H{\sc ii} region with a radius of 1.5\,pc is estimated to be $\sim$0.4\,Myr. 
This is almost consistent with the age of the time scale of the CCC estimated above \textcolor{black}{0.2--0.3\,Myr}, which supports the CCC scenario that explains the formation of the massive star(s). 

\subsection{Massive stars in N4 and N4W}\label{sec:Dis_star}
\begin{figure}[h]
 \begin{center}
   \includegraphics[width=9cm]{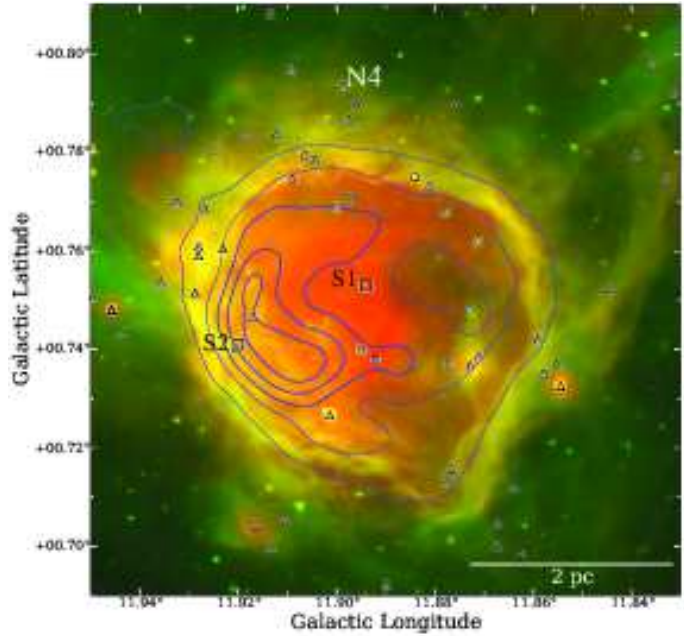}
 \end{center}
  \caption{Closeup figure of Figure\,\ref{fig:RGB_all}, but the blue contours show the intensity of 20-cm radio continuum taken from the Multi-Array Galactic Plane Imaging Survey (MAGPIS, \cite{2006AJ....131.2525H}) archive ($b<0\fdg 8$), and are plotted by 0.75 ($\sim 5\sigma$), 1.95, 3.15, 4.35, and 5.55 mJy\,Beam$^{-1}$. Triangle symbols represent the YSOs identified by \citet{2016ApJ...818...95L}. }\label{fig:RGB_closeup}
\end{figure}

Figure\,\ref{fig:RGB_closeup} shows a closeup of Figure\,\ref{fig:RGB_all}, where the blue contours indicate the intensity of the 20-cm radio continuum taken from the Multi-Array Galactic Plane Imaging Survey (MAGPIS, \cite{2006AJ....131.2525H}) archive. 
As mentioned above, \citet{2013RAA....13..921L} suggested that the \textcolor{black}{most massive and luminous} star in N4 is S1, which is located at the center of the ring, and that the formation of star S2, located on the ring, was triggered by the expansion of the H{\sc ii} region. 
However, the brightest massive stars in the {\it Spitzer} bubbles are not necessarily located at the center (e.g., \cite{2015ApJ...806....7T, 2018PASJ...70S..45O}). 
In N4, \citet{2013RAA....13..921L} showed that the brightest massive star candidate in the $J$ band and the $J-H$ is S2. 
Figure\,\ref{fig:pv_N4_N4W_1}(b) shows a plot of the median intensity (with the direction perpendicular to the B--B$'$) of {\it Spitzer}/MIPSGAL 24 $\mu$m (red) and {\it Spitzer}/GLIMPSE 8$\mu$m (green) along the B--B$'$ line in Figure\,\ref{fig:integ_overlay}(a) with a width of 1.2$'$. 
Of all massive star candidates reported by \citet{2013RAA....13..921L}, the closest to the peak of the 20-cm radio continuum emission, which traces H{\sc ii} regions, is S2, as observed in Figures\,\ref{fig:pv_N4_N4W_1}(b) and \ref{fig:RGB_closeup}. 
For these reasons, we speculate that the \textcolor{black}{most massive and luminous} star in N4 is most likely S2\textcolor{black}{, although we can not rule out the possibility that S1 was most influential on the IR bubble morphology}.
Follow-up near-IR and optical observations would reveal the position and spectral type of the massive star(s) in N4. 

On the contrary, in N4W, no massive star candidates were identified, although \citet{2016ApJ...822..114C} identified one Class I and three Class II YSOs in the innermost area from observations of the $J$, $H$, and $Ks$ bands. 
For these YSOs, the authors derived the crude lower limits of their $L_{\rm bol}$ to be $1$--$2\,\times 10^2\,L_{\odot}$, suggesting at least intermediate masses for the YSOs. 
Using AKARI far-IR data (60, 90, and 140\,$\mu$m) (\cite{2007PASJ...59S.389K}), we estimated the far-IR total luminosity of N4 and N4W by employing the method of \citet{2016A&A...592A.155S}.
As a result, $7.9\,\times 10^4\,L_{\odot}$ and $3.7\,\times 10^4\,L_{\odot}$ were derived for N4 and N4W, respectively.
Since these values correspond to approximately a single O8V star and a single O9.5V star, respectively (\cite{2005A&A...436.1049M}), there might be an embedded massive star in N4W, which has not yet been identified. 

\begin{figure}[h]
 \begin{center}
   \plotone{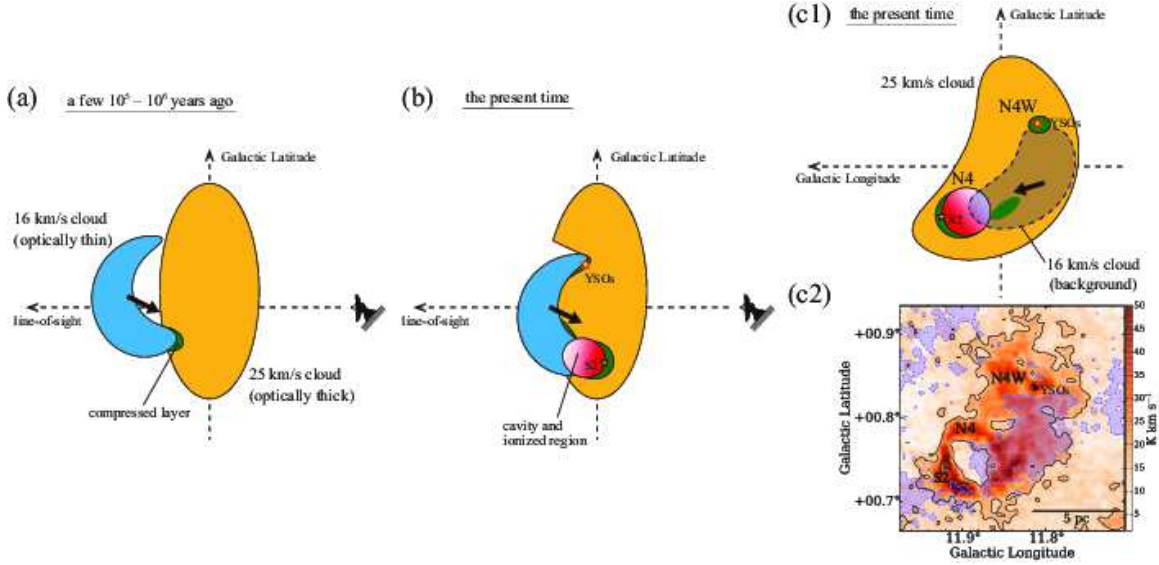}
 \end{center}
  \caption{(a) Sketched diagrams of N4 and N4W as viewed from the Galactic \textcolor{black}{east} when the collision started. Blue and orange show the 16-km\,s$^{-1}$ cloud and the 25-km\,s$^{-1}$ cloud, respectively. Green region indicates the compressed layer between the two clouds, where star(s) form. (b) The sketched diagrams of N4 and N4W as viewed from the Galactic east at the present. Red indicates ionized region. (c1) The sketched diagrams of N4 and N4W on the sky view. (c2) Color scale shows the $^{13}$CO ($J$=1--0) integrated intensity of the 25-km\,s$^{-1}$ cloud (15\,K\,km\,s$^{-1}$). Blue contours shows the $^{12}$CO ($J$=1--0) integrated intensity of the 16-km\,s$^{-1}$ cloud (12\,K\,km\,s$^{-1}$). }\label{fig:sche}
\end{figure}

Figure\,\ref{fig:sche} shows a summary of the assumed CCC scenario discussed in Subsection\,\ref{sec:Dis_ccc}, which considers the location of the massive star in N4 and the YSOs in N4W.
Figure\,\ref{fig:sche}(a) shows a schematic view from the Galactic \textcolor{black}{east} at the time the collision started (a few $10^5$--$10^6$ years ago) and when the molecular clouds in N4 began to be compressed. 
As a result, the 16-km\,s$^{-1}$ cloud created a cavity (the ring-like structure) in the 25-km\,s$^{-1}$ cloud \textcolor{black}{as demonstrated by the numerical simulation of CCC (see Figure\,6 in \citet{2014ApJ...792...63T}).
Thereafter} the massive star S2 was formed in the compressed layer according to the model proposed in a study of \citet{1992PASJ...44..203H}. 
Figure\,\ref{fig:sche}(b) shows a schematic view from the Galactic \textcolor{black}{east} at the present time. 
S2 has either ionized or \textcolor{black}{broken up} the surrounding neutral materials. 
The ionized gas fills up the cavity and erodes the inner surface of the cavity, which is the ring structure observed at 8\,$\mu$m. 
This CCC scenario is able to explain the formation of both the massive star and the ring in N4. 
In addition, the molecular clouds in N4W began to be compressed and YSOs were formed. 
\textcolor{black}{Note that, the curve of the 16-km\,s$^{-1}$ cloud in this sketched diagrams Figure\,\ref{fig:sche}(a) and \ref{fig:sche}(b) is one instance to explain the star formation history in both N4 and N4W.} 
Figures\,\ref{fig:sche}(c1) and \ref{fig:sche}(c2) show schematic views of the sky plane and integrated intensity map of the CO gas, respectively.

\subsection{Comparison of the molecular clouds and massive stars in N4 with those of the other massive-star forming regions}\label{sec:Dis_com}
In Table\,\ref{tab:comparison}, we compare the properties of the colliding molecular clouds in N4 with those of the other massive star-forming regions of RCW 120, RCW 38, and W51A that are also suggested to feature CCC. 
{\it Spitzer} bubble RCW 120 is comparable to N4 in terms of size and its ring appearance in near-IR. 
RCW 38 is known as a super star cluster. 
\citet{2016ApJ...820...26F} suggested that the formation of multiple O-stars was triggered at the point of collision of two clouds. 
The spatial distributions of these two colliding clouds resemble those of the 16-km\,s$^{-1}$ cloud and the 25-km\,s$^{-1}$ cloud in N4. 
W51A is the one of the most active star-forming region in the Galaxy. 
Multiple previous studies (e.g., \cite{1998AJ....116.1856C, 2001PASJ...53..793O, 2017arXiv171101695F}) suggested that a number of velocity components in W51A have been continuously colliding with each other, resulting in active massive star formation. 

\citet{2018ApJ...859..166F} suggested that the molecular column density [$N({\rm H_2})$] of the colliding clouds can be an important parameter for determining the number of the end produce of O-stars.
$N({\rm H_2})$ and the number of O-stars between N4 and RCW 120 were approximately the same, whereas the mass of the associated molecular clouds of RCW 120 were larger by a factor of $\sim$3.
On the contrary, the number of O-stars in RCW 38 is much larger ($\sim$20), although the mass of the associated molecular clouds of RCW 38 is approximately same as that of N4. 
This could be attributed to the higher $N({\rm H_2})$ in RCW 38. 
In W51A, multiple collisions of clouds with high $N({\rm H_2})$ resulted in active massive star formation. 
The relationship between the $V_{\rm LSR}$ separations and the number of O-stars can not be discussed from these data.
To establish a quantitative scenario for forming massive stars via a CCC, more observational studies and statistical studies are required.


\begin{deluxetable*}{ccccccc}
\tablenum{1}
\tablecaption{Properties of the colliding molecular clouds associated with N4, RCW120, RCW 38, and W51A \label{tab:comparison}}
\tablewidth{0pt}
\tablehead{
\colhead{Name} & \colhead{Number of O-stars} & \colhead{Cloud Mass}                 & \colhead{Typical $N({\rm H_2})$}      & $V_{\rm LSR}$ Separation & Age of H{\sc ii} region & References \\
\colhead{}          & \colhead{}                               & \colhead{($10^4\ M_{\odot}$)} & \colhead{($10^{22}\ {\rm cm^{-2}}$)} & \colhead{(km\,s$^{-1}$)}     & \colhead{(Myr)}            & \colhead{}
}
\startdata
N4 & $\sim$1 (O8.5--O9V$^\dagger$) & (1.7, 0.1) & (3--4, 0.3) & 9 & $\sim$0.4 & This study\\
RCW 120 & 1 (O8--O9V) & (5.0, 0.4) & (3, 0.8) & 20 & $\sim$0.2 & \cite{2015ApJ...806....7T}\\
RCW 38 & $\sim$20 O-stars & (2.0, 0.3) & (10, 1) & 12 & $\sim$0.1 & \cite{2016ApJ...820...26F}\\
W51A & $\sim$30 O-stars & (11, 13, 19, 13) & $\sim$10 each & 6--18 & several 0.1 & \cite{2017arXiv171101695F}\\
\enddata
\tablecomments{$\dagger$ \cite{2016ApJ...818...95L}}
\end{deluxetable*}

\section{Summary}\label{sec:Sum}
Using the FUGIN $^{12}$CO ($J$=1--0), $^{13}$CO ($J$=1--0), and the JCMT $^{12}$CO ($J$=3--2) datasets we studied the molecular gas distribution and velocity structure toward the {\it Spitzer} bubble N4 and N4W. The main results and conclusions are summarized below.
\begin{enumerate}
\item We observed three discrete velocity clouds: the 16-, 19-, and 25-km\,s$^{-1}$ clouds. 
Their molecular masses are $0.12 \pm 0.02 \times 10^{4}$\,$M_{\odot}$, $0.43 \pm 0.06 \times 10^{4}$\,$M_{\odot}$, and $3.8 \pm 0.6 \times 10^{4}$\,$M_{\odot}$, respectively. 
The 16- and 19-km\,s$^{-1}$ clouds have not been recognized in previous observations of molecular lines. 
The distribution of the 25-km\,s$^{-1}$ cloud likely traces the ring-like structure observed in the mid-IR wavelength such as 8$\mu$m. 
\item The 16- and 25-km\,s$^{-1}$ clouds are associated with the H{\sc ii} region, whereas the 19-km\,s$^{-1}$ cloud is probably not interacting with N4 and N4W. 
\item We investigated the velocity structure of the molecular clouds associated with the ring in N4, and could not find clear expanding motion. 
\item We found two observational signatures of CCC (bridge features and complementary distributions) between the 16- and 25-km\,s$^{-1}$ clouds in N4. Therefore, we proposed a scenario in which a collision between the two clouds triggered the formation of the massive star candidates in N4 over a short timescale of only $\sim$0.3 Myr. This CCC scenario can explain the formation of both the molecular ring structure and massive star(s) inside the ring. 
\item We also observed a bridge feature between the 16- and 25-km\,s$^{-1}$ clouds also in N4W. The CCC scenario in which massive- or intermediate-mass star formation was triggered by collision between the clouds is therefore also conceivable in N4W. 
\item Both the massive star forming activity and the molecular column density in N4 were comparable to those of RCW120, whose molecular gas distribution also resembles that of N4. The number of O-stars formed via CCC may be related to their molecular column densities rather than to their molecular masses.
\end{enumerate}


\acknowledgments

This study was financially supported by Grants-in-Aid for Scientific Research (KAKENHI) of the Japanese society for the Promotion of Science (JSPS; grant numbers 15K17607, 17H06740, and 18K13580). 
The authors would like to thank the all members of the 45-m group of Nobeyama Radio Observatory for support during the observation. 
Data analysis was carried out on the open use data analysis computer system at the Astronomy Data Center (ADC), of the National Astronomical Observatory of Japan (NAOJ), and made use of  \software{Astropy (\cite{2013A&A...558A..33A}), APLpy (\cite{2012ascl.soft08017R}), astrodendro (\cite{2008ApJ...679.1338R}), SCIMES (\cite{2015MNRAS.454.2067C})}.
The authors also would like to thank NASA, ISAS/JAXA, and Dr. Hong-Li Liu for providing FITS data of {\it Spitzer} Space Telescope, AKARI, and {\it JCMT}, respectively. 
In addition, the authors would like to thank Enago (www.enago.jp) for the English language review.

\appendix

\section{Velocity channel maps of the $^{13}$CO ($J$=1--0) emissions}

\begin{figure}[h]
 \begin{center}
   \includegraphics[width=18cm]{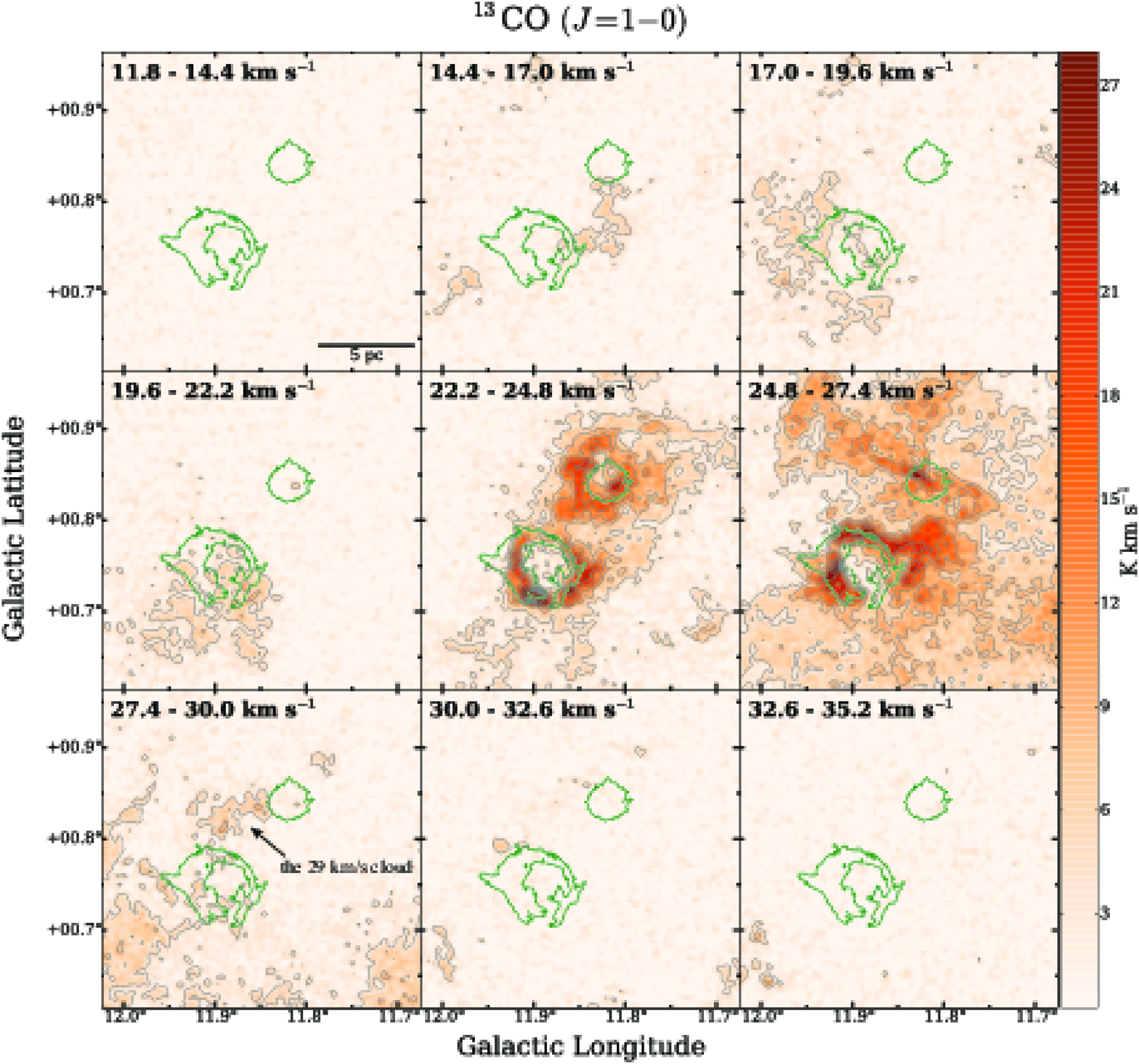}
 \end{center}
  \caption{Same as Figure\,\ref{fig:12COchmap} but for the $^{13}$CO ($J$=1--0) emission. Black contours are plotted at every 4.0\,K\,km\,s$^{-1}$ from 4.0\,K\,km\,s$^{-1}$ ($\sim 5\sigma$). }\label{fig:13COchmap}
\end{figure}

Figure\,\ref{fig:13COchmap} shows the large-scale velocity channel maps of the $^{13}$CO ($J$=1--0) emissions at a velocity step size of 2.6\,km\,s$^{-1}$.

\end{document}